\documentclass[pre,aps,onecolumn,superscriptaddress,nofootinbib]{revtex4-1}
\usepackage{amsmath, amsthm, amssymb,float}
\usepackage{amsfonts}
\usepackage{graphicx}
\usepackage{dcolumn}
\usepackage{bm}
\usepackage{textcomp}
\usepackage[normalem]{ulem}

\usepackage[titletoc,title]{appendix}

\usepackage{color}

\usepackage{colordvi}
\usepackage[usenames,dvipsnames]{xcolor}

\usepackage{epsfig}
\usepackage{textcomp}
\usepackage{epstopdf}

\usepackage[colorlinks=true, urlcolor=blue, anchorcolor=blue, citecolor=blue,filecolor=blue,linkcolor=blue,menucolor=blue
]{hyperref}

\begin{document}

\title{Nonequilibrium Steady State of a Weakly-Driven Kardar-Parisi-Zhang Equation}

\author{Baruch Meerson} \email{meerson@mail.huji.ac.il}
\affiliation{Racah Institute of Physics, Hebrew
  University of Jerusalem, Jerusalem 91904, Israel}

\author{Pavel V. Sasorov} \email{sasorov@gmail.com}
\affiliation{Keldysh Institute of Applied Mathematics, Moscow 125047, Russia}

\author{Arkady Vilenkin} \email{vilenkin@mail.huji.ac.il}
\affiliation{Racah Institute of Physics, Hebrew University of Jerusalem, Jerusalem 91904, Israel}

\begin{abstract}
We consider an infinite interface in $d>2$ dimensions, governed by the
Kardar-Parisi-Zhang (KPZ) equation with a weak Gaussian noise which
is delta-correlated in time and has short-range spatial correlations.
We study the probability distribution of the interface height $H$ at a point of the substrate, when the interface is initially flat. We show that, in a stark contrast with the KPZ equation in $d<2$,  this distribution approaches a non-equilibrium steady state. The time of relaxation toward this state
scales as the diffusion time over the correlation length of the noise.
We study the steady-state distribution $\mathcal{P}(H)$ using the optimal-fluctuation method.  The typical, small fluctuations of height are Gaussian. For these fluctuations the activation path of the system coincides with the time-reversed relaxation path, and the variance of $\mathcal{P}(H)$  can be found from a minimization of the (nonlocal) equilibrium free energy of the interface. In contrast, the tails of $\mathcal{P}(H)$ are nonequilibrium, non-Gaussian and strongly asymmetric. To determine them we calculate, analytically and numerically,  the activation paths of the system, which are different from the time-reversed relaxation paths. We show that the slower-decaying tail of $\mathcal{P}(H)$ scales
as $-\ln \mathcal{P}(H) \propto |H|$, while the faster-decaying tail scales as $-\ln \mathcal{P}(H) \propto |H|^3$. The slower-decaying tail has important implications for the statistics of directed polymers in random potential.
\end{abstract}

\maketitle

\tableofcontents
\nopagebreak

\section{Introduction}

For the last 30 years the Kardar-Parisi-Zhang (KPZ) equation \cite{KPZ}
has been in the focus of attention of theoretical and experimental physicists
and mathematicians \cite{Krugbook,Barabasi,HHZ,Krug,Corwin,Hairer,QS,HHT,S2016,Sasamotoreview,Kupiainen,Takeuchi}
and has become a paradigmatic model of
nonequilibrium statistical mechanics. This equation provides a continuum description of kinetic
roughening of growing stochastic surfaces, and it has been remarkably successful in capturing
universal scaling properties of a whole class of microscopic models \cite{Corwin,QS,S2016,Sasamotoreview,Takeuchi}
and, no less important, of some real systems \cite{Barabasi,Takeuchi}. The KPZ equation,
\begin{equation}\label{KPZda}
\partial_{t}h=\nu \nabla^2 h+\frac{\lambda}{2}\left(\nabla h\right)^2+\sqrt{D}\,\xi(\mathbf{x},t),
\end{equation}
describes the evolution of a fluctuating surface height $h(\mathbf{x},t)$,
where $\mathbf{x}$ is a $d$-dimensional position vector in the substrate hyperplane of
growth. The homogeneous and isotropic Gaussian noise $\xi(\mathbf{x},t)$ has zero mean and
the correlator
\begin{equation}\label{CORRELATOR}
\langle\xi(\mathbf{x}_{1},t_{1})\xi(\mathbf{x}_{2},t_{2})\rangle = \kappa(\mathbf{x}_{1}-\mathbf{x}_{2})\,\delta(t_{1}-t_{2}),
\end{equation}
where $\kappa(\mathbf{x})$ can be normalized to unity:
\begin{equation}\label{normcorr}
\int d \mathbf{x} \,\kappa(\mathbf{x}) =1 .
\end{equation}
In their original paper \cite{KPZ} the authors assumed that the noise is delta-correlated in space, $\kappa(\mathbf{x})=\delta(\mathbf{x})$. Later it became clear that, in this case, Eq.~(\ref{KPZda}) is ill-defined. Indeed, already at $d=1$ the systematic velocity of the interface,
which results from the rectification of the noise by the nonlinear term, diverges unless the noise has a finite spatial correlation length $\ell>0$ \cite{S2016}. The latter can be formally defined as
\begin{equation}\label{corrlengthdefine}
\ell^2 = \int d \mathbf{x} \,\mathbf{x}^2 \kappa(\mathbf{x}).
\end{equation}
After subtracting the systematic interface shift, many interesting
statistical quantities  become well-defined, at $d=1$, even for $\ell=0$.
However, at $d\geq 2$ even the properly shifted quantities, such as the variance of the fluctuating interface width \cite{Krug,NattermannTang}, or the variance of the one-point height distribution \cite{KK2008,SMS}, diverge unless $\ell$ is finite. Being interested in $d>2$,  we will assume throughout this work a finite correlation length of the noise.

Since the inception of the KPZ equation, the central questions in its analysis have been about statistical self-affine properties and dynamic scaling of the interface \cite{KPZ,Krugbook,Barabasi}. For $d=1$ the KPZ-interface exhibits roughening which, in an infinite system, would continue forever for any nonzero noise strength.
More recently, the focus of studies in $1d$ has shifted toward more detailed characterizations of the roughening interface. One of them is the  one-point probability distribution $\mathcal{P}_t(H)$ of the (properly shifted) surface height $H$ at a specified time $t$ \cite{Corwin,QS,HHT,S2016,Sasamotoreview,Takeuchi}.  A remarkable analytic progress has been achieved, in the form of exact representations for $\mathcal{P}_t(H)$, for several initial conditions \cite{Corwin,QS,S2016} and some of their combinations for an infinite system. In this case $\mathcal{P}_t(H)$ is time-dependent, and quite sensitive to the initial conditions, even at late times.

For $d>1$ analytic progress has been slow and painful, and exact results are unavailable. Still, it is known that,  at $d>2$, there are two
distinct regimes: of weak and strong noise, also called weak and strong coupling.  For weak noise, an infinite KPZ interface becomes smooth at long times. For strong noise it continues to roughen forever. A nearly universal consensus is that the transition between the two regimes [as a function of the dimensionless noise magnitude $\epsilon$, introduced in Eq.~(\ref{epsilon}) below] has a character of a phase transition, whereas $d=2$ is a critical dimension \cite{Barabasi,HHZ}.

Understandably, most of the numerical and analytical efforts at $d\geq 2$ are being spent on the strong-coupling regime.  Here we consider the weak-coupling regime at $d>2$ which turns out to be much simpler for analysis, and where we can achieve a good quantitative understanding of the \emph{complete statistics} of the one-point height fluctuations. We argue that the KPZ interface, driven by a weak noise, exhibits very interesting properties which have been overlooked. We show that even an infinite interface relaxes to a well-defined \emph{nonequilibrium steady state}, so that the (properly shifted) one-point probability distribution becomes time-independent, $\mathcal{P}_t(H)=\mathcal{P}(H)$.  For an initially flat interface, the characteristic time that it takes the typical fluctuations to reach the stationary distribution scales as the diffusion time over the correlation length scale. When the noise correlations come from a microscopic regularization of the KPZ equation,  this time is very short.

We study this stationary distribution $\mathcal{P}(H)$ by the optimal-fluctuation method (OFM), also known as the instanton method \cite{FKLM,GM,BFKL,CS,GGS}, the weak-noise theory \cite{F1998,F1999,F2009}, the macroscopic fluctuation theory \cite{MFTreview}, \textit{etc}. The crux of the OFM is a saddle-point evaluation of the path integral for the KPZ equation~(\ref{KPZda}), constrained by the specified large deviation. The ensuing minimization procedure generates an effective classical field theory. The solution of the minimization problem yields the ``optimal fluctuation" (also known as optimal path, activation path, or optimal history). The optimal fluctuation dominates the contribution to the probability
of observing a specified height $H$ at a specified point [which can be set to be $\mathbf{x}=0$]. The minus logarithm of the probability distribution is approximately equal to the classical action along the optimal path. The OFM has been extensively used for studying $\mathcal{P}_t(H)$ of the KPZ equation for different initial conditions in one dimension \cite{KK2007,KK2008,KK2009,MKV,KMS,JKM,MSchmidt,SMS_KPZ,SKM,SM}. In higher dimensions, some important results were obtained by Kolokolov and Korshunov \cite{KK2008,KK2009}, but the stationarity of $\mathcal{P}(H)$ at $d>2$ was not appreciated.

A natural rescaling of the weakly-driven KPZ equation at $d>2$  is the following. We rescale $\mathbf{x}$ by the correlation length of the noise $\ell$, $t$ by the diffusion length $\ell^2/\nu$, and $h$ by $\nu/|\lambda|$. This brings Eq.~(\ref{KPZda}) to the rescaled form\footnote{\label{flip} Without losing generality we assume that $\lambda<0$.  Flipping the sign of $\lambda$ is equivalent to flipping the sign of $h$.}
\begin{equation}\label{KPZrescaled}
\partial_{t}h=\nabla^2 h-\frac{1}{2}\left(\nabla h\right)^2+\sqrt{\epsilon} \,\xi(\mathbf{x},t),
\end{equation}
where
\begin{equation}\label{epsilon}
\epsilon=\frac{\lambda^2 D}{\nu^3 \ell^{d-2}}
\end{equation}
is the dimensionless noise strength. Equation~(\ref{CORRELATOR}) keeps the same form in the rescaled variables.  For an initially flat interface, the long-time evolution of the KPZ-interface is determined solely by the effective noise strength $\epsilon$. Importantly,  $\epsilon$ does not depend on the observation time $t$. This is in contrast with $d=1$ \cite{MKV}, where the properly defined effective noise strength scales as $\sqrt{t}$ \cite{KK2007,MKV} and, as a result, the noise is always strong at long times.

Here are the main findings of this work. A key observation is that, for $d>2$ and at small $\epsilon$, the whole probability distribution  $\mathcal{P}_t(H)$ approaches a steady state. This property is closely related to the fact that the dominant contribution to the local fluctuations of the height, for weak coupling, comes from the noise which spatial scale is comparable with the correlation length \cite{KK2008}\footnote{\label{early} When the nonlinear term in Eq.~(\ref{KPZrescaled}) is dropped, the statistical stationarity of the interface at $d>2$ at long times is evident in the behavior of the mean-square width of the interface and in the height-height correlation function \cite{NattermannTang}. Here we emphasize the stationarity of the whole one-point height distribution $\mathcal{P}(H)$, including its non-Gaussian tails, in the weak-coupling regime of the KPZ equation.}.
Furthermore, in the weak-noise regime, $\epsilon \ll 1$, $ \ln \mathcal{P}(H)$ exhibits the following scaling behavior:
\begin{equation}\label{ansatz}
-\ln \mathcal{P}(H)\simeq \frac{s(\tilde{H})}{\epsilon} = \frac{\nu^3 \ell^{d-2}}{\lambda^2 D}\,s\left(\frac{|\lambda| H}{\nu}\right),
\end{equation}
where the  large-deviation function $s(\tilde{H})$ depends on $d$ and on the particular form of the spatial correlator of the noise  $\kappa(\mathbf{x})$. The asymptotic behavior of $s(\tilde{H})$ is the following. For small $H$ the function $s(\dots)$ is a quadratic function of its argument, as to be expected. This regime is described by the Edwards-Wilkinson equation. Correspondingly, the variance of  $\mathcal{P}(H)$  is independent of $\lambda$ and equal to
\begin{equation}\label{variance}
v_H=\frac{C_0 D}{\nu \ell^{d-2}} ,\quad d>2\,,
\end{equation}
where the non-universal constant $C_0$ depends on $d$ and on the particular form of $\kappa(\mathbf{x})$. As we show here, Eq.~(\ref{variance}) can be obtained via a minimization of a (nonlocal) ``equilibrium free energy" of the interface. Here there is no need to deal with the interface dynamics for the purpose of calculation of $v_H$.

The more interesting asymptotics concern \emph{large deviations} of the interface height which are intrinsically nonequilibrium. These are described by the (non-Gaussian and strongly asymmetric) tails of $\mathcal{P}(H)$. To determine the tails we compute, analytically and numerically, the corresponding optimal paths, which differ from the time-reversed relaxation paths.
As we show, the slower-decaying tail $\lambda H>0$ scales, in the original variables,  as
\begin{equation}\label{slowtail}
-\ln \mathcal{P}(H) \simeq \frac{C_s \nu^2 \ell^{d-2} |H|}{|\lambda| D}\, ,\quad d>2\,,
\end{equation}
while the faster-decaying tail $\lambda H<0$ behaves as
\begin{equation}\label{fasttail}
-\ln \mathcal{P}(H)\simeq \frac{C_f |\lambda| \ell^{d-2} |H|^3}{D}\, ,\quad d>2\,,
\end{equation}
and is independent of $\nu$.
The constants $C_s$ and $C_f$ depend on $d$ and on the particular form of $\kappa(\mathbf{x})$. We show how to compute these constants numerically: by using the Chernykh-Stepanov back-and-forth iteration algorithm \cite{CS} for both tails, or by solving a nonlinear integro-differential equation [Eq.~(\ref{H234}) below] for the slower-decaying tail~(\ref{slowtail}).

Here is how we organized the remainder of the paper. In Sec.~\ref{OFM} we briefly introduce the OFM and present its governing equations and the boundary conditions. Section~\ref{EW} deals with
\emph{typical} fluctuations of the weakly-driven KPZ interface, where the KPZ nonlinearity can be neglected. We derive the free energy of the interface in this limit and determine the most likely interface shape conditioned on reaching a specified height at a point. In Sec.~\ref{negative} we evaluate the slower-decaying, $\lambda H>0$, tail of $\mathcal{P}(H)$ by analyzing static soliton solutions of the OFM equations and solving an ensuing selection problem. Section~\ref{positive} deals with the faster-decaying, $\lambda H<0$, tail of $\mathcal{P}(H)$ by solving the OFM equations analytically in the inviscid limit $\nu \to 0$. The results of Sections~\ref{EW}, \ref{negative} and \ref{positive} are supported by numerical solutions of the OFM equations. The slower-decaying tail behavior (\ref{slowtail}) has important implications for the statistics of the partition function of directed polymers in random potential. We discuss them in Section~\ref{polymer}. Our results are summarized and discussed in Sec.~\ref{SD}. A brief description of two numerical methods that we used are relegated to the Appendix.

\section{Optimal-fluctuation method: governing equations}
\label{OFM}

The OFM employs the small parameter $\epsilon$, see Eq.~(\ref{epsilon}), for a saddle-point evaluation of the properly constrained path integral of Eq.~(\ref{KPZrescaled}). Let us  introduce the kernel $K(\mathbf{x}-\mathbf{x}^\prime)$ which is inverse to the correlation kernel $\kappa(\mathbf{x}-\mathbf{x}^\prime)$:
\begin{equation}\label{inverse}
\int d\mathbf{x}^{\prime\prime} \,K(\mathbf{x}-\mathbf{x}^{\prime\prime}) \,\kappa(\mathbf{x}^{\prime}-\mathbf{x}^{\prime\prime}) =\delta(\mathbf{x}-\mathbf{x}^{\prime}).
\end{equation}
The saddle-point evaluation procedure leads to a minimization of the effective ``classical action"
\begin{equation}\label{Func}
{\cal S}[h(\mathbf{x},t)]=\frac{1}{2}\int_{-\infty}^0\, dt
\int d\mathbf{x}\, d\mathbf{x}^\prime
\left(\partial_th-\nabla^2h+\frac{1}{2}(\nabla h)^2\right)\biggr|_{(\mathbf{x},t)}
\, K\left(\mathbf{x}-\mathbf{x}^\prime\right)\,
\left(\partial_th-\nabla^2h+\frac{1}{2}(\nabla h)^2\right)\biggr|_{(\mathbf{x}^\prime,t)}\,.
\end{equation}
The ensuing Euler-Lagrange equation can be cast into a Hamiltonian form.  For a spatially-correlated and temporally uncorrelated Gaussian noise this procedure, by now fairly standard, was performed by Gurarie and Migdal \cite{GM} for the noise-driven Burgers equation, and by Kolokolov and Korshunov \cite{KK2008} for the KPZ equation itself. Therefore, we can be brief here.  The resulting Hamilton equations take the form
\begin{eqnarray}
\partial_{t}h &=& \nabla^2 h -\frac{1}{2} \left(\nabla h\right)^2+ \pi(\mathbf{x},t)\,,\label{heqd}\\
  \partial_{t}\rho &=& -\nabla^2 \rho -\nabla \cdot \left(\rho \nabla h\right)\,. \label{peqd}
\end{eqnarray}
Here the optimal history of the interface $h(\mathbf{x},t)$ plays the role of the Hamiltonian coordinate, while the optimal realization of the noise $\rho(\mathbf{x},t)$ plays the role of the conjugate momentum. Further,
\begin{equation}\label{pi}
\pi(\mathbf{x},t)= \int d\mathbf{x}^{\prime} \kappa(\mathbf{x}-\mathbf{x}^{\prime})  \rho(\mathbf{x}^{\prime},t)
\end{equation}
is an auxiliary field which is related to $\rho(\mathbf{x},t)$ nonlocally. The nonlocality is a consequence of spatial correlations of the noise.  The Hamiltonian of Eqs.~(\ref{heqd}) and (\ref{peqd}) is
$H=\int d\mathbf{x} \,\mathcal{H}$, where
\begin{equation}\label{hamdens}
 \mathcal{H}(\mathbf{x},t)= \rho \left[\nabla^2 h-\frac{1}{2} \left(\nabla h\right)^2 +\frac{1}{2}\pi(\mathbf{x},t)\right]\,.
\end{equation}
Equations~(\ref{heqd}) and (\ref{peqd}) must be supplemented by boundary conditions in space and in time.  Being interesting in a steady-state distribution, we can assume a noiseless flat interface at $t=-\infty$:
\begin{equation}\label{timeminusinfinity}
h(\mathbf{x}, t=-\infty)= \rho(\mathbf{x}, t=-\infty) = 0.
\end{equation}
The interface must reach a specified rescaled height $H$ (we omit the tildes) at $\mathbf{x}=0$ and $t=0$:
\begin{equation}\label{htimezero}
h(\mathbf{x}=0, t=0)= H.
\end{equation}
This condition leads to a singular boundary condition on $\rho(\mathbf{x},t)$ \cite{KK2008}:
\begin{equation}\label{ptimezero}
\rho(\mathbf{x}=0, t=0)= \Lambda \, \delta (\mathbf{x}),
\end{equation}
where the Lagrangian multiplier $\Lambda$ is ultimately set by the value of $H$ from Eq.~(\ref{htimezero}). Finally, both $h(\mathbf{x},t)$ and $\rho(\mathbf{x},t)$ must vanish at $\mathbf{x} \to \infty$. The problem possesses the conservation law
\begin{equation}\label{conservation}
\int d \mathbf{x} \,\rho(\mathbf{x},t) = \Lambda .
\end{equation}
Once the optimal path, conditioned by Eq.~(\ref{htimezero}), is found, $\mathcal {P}(H)$ is given in terms of the classical action~(\ref{Func}), evaluated along this path:
\begin{equation}
-\ln \mathcal {P}(H) \simeq  \frac{1}{\epsilon} \int_{-\infty}^0 dt \int d\mathbf{x} \left(\rho \partial_t h - \mathcal{H} \right) = \frac{s(H)}{\epsilon},
\end{equation}
where
\begin{equation}\label{sd}
s(H) = \frac{1}{2}\int_{-\infty}^0 dt \int d\mathbf{x} \, \rho(\mathbf{x},t)\,\pi(\mathbf{x},t)=\frac{1}{2}\int_{-\infty}^0 dt \int d\mathbf{x}  \int d\mathbf{x}^{\prime} \rho(\mathbf{x},t) \kappa(\mathbf{x}-\mathbf{x}^{\prime}) \,\rho(\mathbf{x}^{\prime},t)\,,
\end{equation}
see Ref. \cite{KK2008}. Let us also define the action \emph{accumulation rate} on the optimal path, which we will call $\dot{s}(H,t)$:
\begin{equation}\label{sdotdef}
\dot{s}(H,t) = \frac{1}{2}\int d\mathbf{x} \, \rho(\mathbf{x},t)\,\pi(\mathbf{x},t) .
\end{equation}
Using the inverse kernel $K$, we can rewrite Eq.~(\ref{sd}) as
\begin{equation}\label{sdpi}
s(H)= \frac{1}{2}\int_{-\infty}^0 dt \int d\mathbf{x}  \int d\mathbf{x}^{\prime} \pi(\mathbf{x},t) K(\mathbf{x}-\mathbf{x}^{\prime}) \,\pi(\mathbf{x}^{\prime},t).
\end{equation}
In the following we will need to use the Fourier transforms of $\kappa(\boldsymbol{\xi})$ and $K(\boldsymbol{\xi})$:
\begin{equation}\label{Fourier}
\kappa_{\mathbf{k}} = (2\pi)^{-d} \int d \boldsymbol{\xi} \,e^{-i \mathbf{k} \boldsymbol{\xi}} \kappa(\boldsymbol{\xi})
\quad \mbox{and} \quad K_{\mathbf{k}} = (2\pi)^{-d} \int d \boldsymbol{\xi} \,e^{-i \mathbf{k} \boldsymbol{\xi}} K(\boldsymbol{\xi}) .
\end{equation}
They are related as follows:
\begin{equation}\label{kappaK}
\kappa_{\mathbf{k}} K_{\mathbf{k}} = (2\pi)^{-2d}.
\end{equation}

\section{Typical fluctuations and equilibrium free energy}
\label{EW}

For small fluctuations we can drop the nonlinear terms in Eqs.~(\ref{heqd}) and (\ref{peqd}) and arrive at
the linear equations
\begin{eqnarray}
  \partial_{t}h &=& \nabla^2 h+\pi(\mathbf{x},t),  \label{heqlin}\\
  \partial_t \rho &=& -\nabla^2 \rho, \label{peqlin}
\end{eqnarray}
which describe typical, small fluctuations of the interface. For the Edwards-Wilkinson interface these equations would be exact.

A straightforward way of solving the linear problem in the framework of the OFM would be the following. One first solves the anti-diffusion equation~(\ref{peqlin}) backward in time with the ``initial" condition~(\ref{ptimezero}). Then one evaluates $\pi(\mathbf{x},t)$, using Eq.~(\ref{pi}), and solves the driven diffusion equation~(\ref{heqlin}) forward in time. The action is then calculated from Eq.~(\ref{sd}), and $\Lambda$ is expressed through $H$ via Eq.~(\ref{htimezero}). The following shortcut simplifies the calculations. As the linearized system is in equilibrium,
the activation path $h(\mathbf{x},t)$, defined on the time interval $-\infty<t\leq 0$, must coincide with the \emph{time-reversed} relaxation path  $h_r(\mathbf{x},t)$, defined on the interval $0\leq t<+\infty$ \cite{Onsager}. The momentum field of the relaxation path is zero: $\rho_r(\mathbf{x},t)=0$, $0\leq t<+\infty$. As a result, the activation path $h(\mathbf{x},t)$ obeys the equation
\begin{equation}\label{timereversed}
 \partial_{t}h(\mathbf{x},t) = -\nabla^2 h(\mathbf{x},t) .
\end{equation}
Combining Eqs.~(\ref{heqlin})-(\ref{timereversed}), we obtain two equilibrium relations,
\begin{equation}\label{equil}
\pi(\mathbf{x},t)=-2 \nabla^2 h(\mathbf{x},t) \quad\mbox{and}\quad \pi(\mathbf{x},t)=2 \partial_t h(\mathbf{x},t).
\end{equation}
Once Eq.~(\ref{peqlin}) for $\rho(\mathbf{x},t)$ is solved, the second relation in Eq.~(\ref{equil}) in conjunction with Eq.~(\ref{pi}) allows one to compute $h(\mathbf{x},t)$ by  integration over time.

There is, however, a more radical shortcut which makes the dynamic calculations altogether redundant, as is indeed to be expected from an equilibrium system (\ref{heqlin}) and (\ref{peqlin}).
Using the relations~(\ref{equil}) directly in Eq.~(\ref{sdpi}) for the action, and performing integrations by parts, we obtain
\begin{equation}\label{freeenergy}
s\left[\phi(\mathbf{x})\right]=
\int  d\mathbf{x}\int d\mathbf{x}^\prime K(\mathbf{x}-\mathbf{x}^\prime)\, \nabla \phi(\mathbf{x})\cdot
\nabla \phi(\mathbf{x}^\prime) ,
\end{equation}
where $\phi(\mathbf{x})\equiv h(\mathbf{x},t=0)$ is the optimal shape of the interface at the observation time $t=0$. Equation~(\ref{freeenergy}) does not include integration over time. It  describes the \emph{equilibrium free energy} of the Edwards-Wilkinson interface with a spatially-correlated noise. Together with the relation $-\ln \mathcal{P} \simeq s(H)/\epsilon$, the free energy provides a complete description of the equilibrium interface, which represents a random Gaussian field.

At $\epsilon \ll 1$ the free energy suffices for computing the variance of $\mathcal{P}(H)$, up to small sub-leading terms, for the KPZ interface.  Let us perform this calculation.
We should minimize the free energy~(\ref{freeenergy}) subject to the boundary conditions $\phi(\mathbf{x}=0)=H$ and  $\phi(\mathbf{x}\to \infty)=0$. It is convenient to recast the boundary condition  $\phi(\mathbf{x}=0)=H$ as an integral constraint:
\begin{equation}\label{HHH}
\int \phi(\mathbf{x}) \delta(\mathbf{x}) d\mathbf{x} =H.
\end{equation}
Now we can minimize the functional
\begin{equation}\label{functional}
\tilde{s}[\phi(\mathbf{x})]= \int d\mathbf{x} \left[\int d\mathbf{x}'  K(\mathbf{x}-\mathbf{x}') \nabla \phi(\mathbf{x}) \cdot \nabla \phi(\mathbf{x}') -\Lambda_0 \phi(\mathbf{x}) \delta(\mathbf{x})\right] ,
\end{equation}
where $\Lambda_0$ is a Lagrange multiplier [not to be confused with the Lagrange multiplier $\Lambda$ entering
Eq.~(\ref{ptimezero})].
Demanding that the variation vanish, we obtain
\begin{equation}\label{B100}
\int d\mathbf{x}'\,K(\mathbf{x}-\mathbf{x}')\nabla^2 \phi(\mathbf{x}') + \frac{\Lambda_0}{2} \delta(\mathbf{x})=0
\end{equation}
or, using Eq.~(\ref{inverse}) and interchanging $\mathbf{x}$ and $\mathbf{x}'$,
\begin{equation}\label{B110}
\int d\mathbf{x} K(\mathbf{x}-\mathbf{x}')\left[\nabla^2 \phi(\mathbf{x}) + \frac{\Lambda_0}{2} \kappa(\mathbf{x})\right]=0.
\end{equation}
As this equality must hold for any $\mathbf{x}'$, we arrive at the Euler-Lagrange equation for the functional $\tilde{s}\left[\phi(\mathbf{x})\right]$:
\begin{equation}\label{EL}
\nabla^2 \phi(\mathbf{x}) = - \frac{\Lambda_0}{2} \kappa(\mathbf{x}) .
\end{equation}
This is a Poisson equation for an effective potential $\phi(\mathbf{x})$. Its Fourier transform is
\begin{equation}\label{phik}
\phi_\mathbf{k}=\frac{\Lambda_0}{2 k^2}\,\kappa_{\mathbf{k}}.
\end{equation}
The knowledge of $\phi_\mathbf{k}$ suffices for calculating $s\left[\phi(\mathbf{x})\right]$ in terms of $\Lambda_0$. Indeed, using Eq.~(\ref{kappaK}), we obtain
$$
K(\mathbf{x}-\mathbf{x}')=\frac{1}{(2 \pi)^{2d}}\int \frac{d\mathbf{k}}{\kappa_{\mathbf{k}}} \,e^{i \mathbf{k}(\mathbf{x}-\mathbf{x}') }
$$
and recast Eq.~(\ref{freeenergy}) as
$$ s\left[\phi(\mathbf{x})\right]=\frac{1}{(2 \pi)^{2d}}\int \frac{d\mathbf{k}}{\kappa_{\mathbf{k}}}\int d\mathbf{x} \,e^{i \mathbf{k}\mathbf{x}}\, \nabla \phi(\mathbf{x}) \cdot
\int d\mathbf{x}^\prime \,e^{-i \mathbf{k}\mathbf{x}'}\nabla \phi(\mathbf{x}')
=\int d\mathbf{k}\,\frac{k^2 \phi_\mathbf{k}^2}{\kappa_{\mathbf{k}}}.
$$
Plugging here $\phi_\mathbf{k}$ from Eq.~(\ref{phik}), we obtain
\begin{equation}\label{svsLambdalinear}
s\left[\phi(\mathbf{x})\right]=\frac{\Lambda_0^2}{4} \int d \mathbf{k} \,\frac{\kappa_{\mathbf{k}}}{k^2}.
\end{equation}
What remains to do is to express $\Lambda_0$ through $H$. For that we need to solve Eq.~(\ref{EL}) in the real space.
As the noise is isotropic, and the solution is unique, we can exploit spherical symmetry
and rewrite Eq.~(\ref{EL}) as an ordinary differential equation (ODE):
\begin{equation}\label{ELeqn}
\frac{1}{r^{d-1}} \frac{d}{dr} \left(r^{d-1}\frac{d\phi}{dr}\right)= -\frac{\Lambda_0}{2} \kappa(r)\,.
\end{equation}
The solution $\phi(r)$, which vanishes at $r=\infty$ and is equal to $H$ at $r=0$, is obtained
by two consecutive integrations.  As a result,
\begin{equation}\label{Lambdalinear}
\Lambda_0=\frac{2H}{\int_0^{\infty} dr \int_0^r dr'\,(r'/r)^{d-1} \kappa(r')} .
\end{equation}
Equations (\ref{svsLambdalinear}) and (\ref{Lambdalinear}) yield the announced variance $v_H$ from Eq.~(\ref{variance}), where
\begin{equation}\label{C0}
C_0=\frac{\Gamma(d/2) \left[\int_0^{\infty} dr \int_0^r dr'\,(r'/r)^{d-1} \kappa(r')\right]^2}
{4 \pi^{d/2}  \int_0^{\infty} dk\, k^{d-3} \kappa_k},
\end{equation}
and $\Gamma(z)$ is the Euler gamma function \cite{Wolfram}.

\begin{figure} [ht]
\includegraphics[width=0.45\textwidth,clip=]{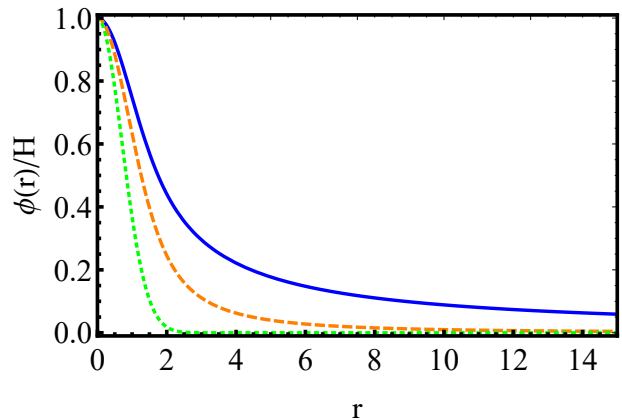}
\caption{The most probable shapes of equilibrium Edwards-Wilkinson interfaces reaching the height $H$ at $d=3, 4$ and $\infty$ (top to bottom) for the spatial correlator~(\ref{correlator}) of the noise. The interface height $\phi(r)$ is rescaled by its value $H$ at $r=0$. The radial distance is measured in units of the correlation length $\ell$ of the noise.}
\label{3phi}
\end{figure}

As a concrete example, here and in the following we consider a Gaussian correlator. In the original variables it is \footnote{\label{factor}Here, for convenience, we slightly changed the definition  of the correlation length $\ell$ compared to Eq.~(\ref{corrlengthdefine}). The new $\ell$ is equal to the old one multiplied by $\sqrt{2/d}$.}
\begin{equation}\label{correlator}
\kappa(z)= \left(\pi \ell^2\right)^{-d/2}\, e^{-z^2/\ell^2} .
\end{equation}
In the rescaled variables $\kappa(z)= \pi^{-d/2}\, e^{-z^2}$.
Then Eq.~(\ref{C0}) yields
\begin{equation}\label{C0example}
C_0=\frac{1}{4 (d-2) \pi^{d/2}} .
\end{equation}
In this example the optimal interface shape at the observation time is
\begin{equation}\label{PhidLambda}
\phi(r) =H \left[e^{-r^2}-\frac{\Gamma \left(\frac{d}{2},r^2\right)-\Gamma
   \left(\frac{d}{2}\right)}{r^{d-2}}\right] ,
\end{equation}
where  $\Gamma(a,z)$ is the incomplete gamma function \cite{Wolfram}.  At large distances $\phi(r)\sim 1/r^{(d-2)}$ is the $d$-dimensional Coulomb potential of a point-like charge placed at the origin.
For $d=3$ and $4$ Eq.~(\ref{PhidLambda}) yields
\begin{equation}\label{3and4}
\phi_3 (r)=\frac{\sqrt{\pi} H \,\text{erf}\,r}{2r}\quad \,\mbox{and}\quad \phi_4 (r)=\frac{H(1-e^{-r^2})}{r^2},
\end{equation}
respectively. Here $\text{erf} \,z$ is the error function \cite{Wolfram}. For $d \to \infty$ we obtain $\phi_{\infty} (r) = H\,e^{-r^2}$: the Coulomb contribution disappears, and only a short-range contribution (on the correlation length scale of the noise) remains. The functions $\phi_3(r), \phi_4(r)$ and $\phi_{\infty}(r)$ are shown in Fig. \ref{3phi}.

\section{The $\lambda H >0$ tail}
\label{negative}

As in the well-studied case of $d=1$ \cite{KK2007,KK2008,MKV}, the $\lambda H >0$ tail is dominated by the KPZ nonlinearity, but diffusion still plays an important role. Its competition with the nonlinearity determines the size of the relatively small region around the origin where $\rho(\mathbf{x},t)$, the optimal fluctuation of the noise field, is localized. At moderately large heights the size of this region is much larger than the correlation length $1$. At still larger heights the inequality is opposite. In all cases, $\rho(\mathbf{x},t) \simeq 0$ outside of this region. Here the optimal path of the system is almost deterministic and can be approximately described by the noiseless KPZ equation
\begin{equation}\label{Hopf1}
\partial_{t}h =\nabla^2 h -\frac{1}{2} \left(\nabla h\right)^2.
\end{equation}

\subsection{$\rho$-solitons and $h$-fronts}

A key role in determining the $\lambda H>0$ tail is played, in all dimensions, by a family of solutions of  Eqs.~(\ref{heqd}) and~(\ref{peqd}) which describe a stationary pulse of the $\rho$-field, which we call soliton, and a traveling front of $h(\mathbf{x},t)$ which this soliton drives \cite{KK2008}:
\begin{equation}\label{ansatz1}
\rho\left(\mathbf{x},t\right)=\mathrm{P}\left(\mathbf{x}\right)\,,\quad
h \left(\mathbf{x},t\right)= \mathrm{H}\left(\mathbf{x}\right)-ct\,, \quad c>0\,.
\end{equation}
These solutions can be parametrized by the velocity $c$ of the traveling $h$-front. A larger $c$ corresponds to a larger amplitude and smaller width of the $\rho$-solitons. At $d=1$ the height distribution never reaches a steady state. As a result, velocity $c$ is uniquely selected by the specified interface height $H$ and the specified time $t$ when this height is reached, and one obtains $c\simeq |H|/t$  \cite{KK2007,MKV}.  At $d>2$,  time (if it is sufficiently large) drops out of the problem, and $c$ must be selected by minimizing the action. This important aspect of the high-dimensional case was not appreciated previously.

\begin{figure} [ht]
\includegraphics[width=0.45\textwidth,clip=]{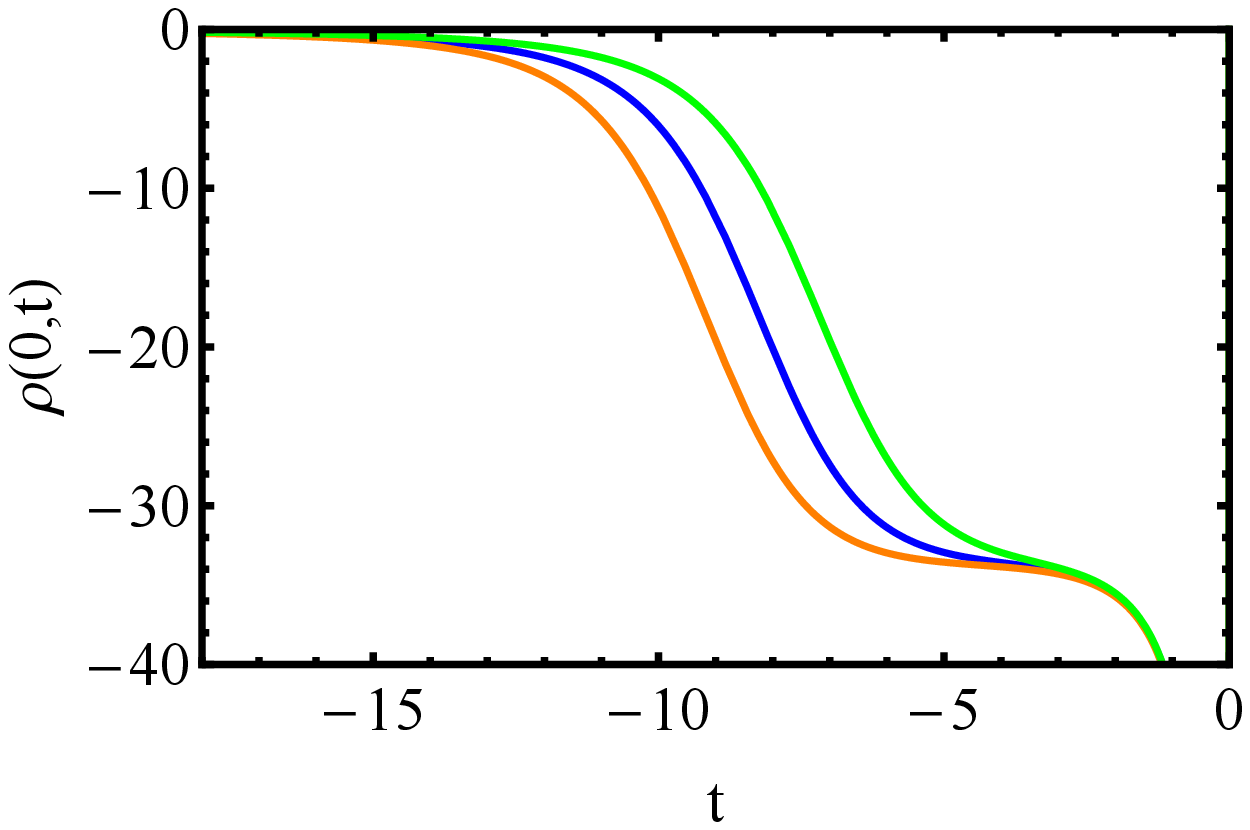}
\includegraphics[width=0.45\textwidth,clip=]{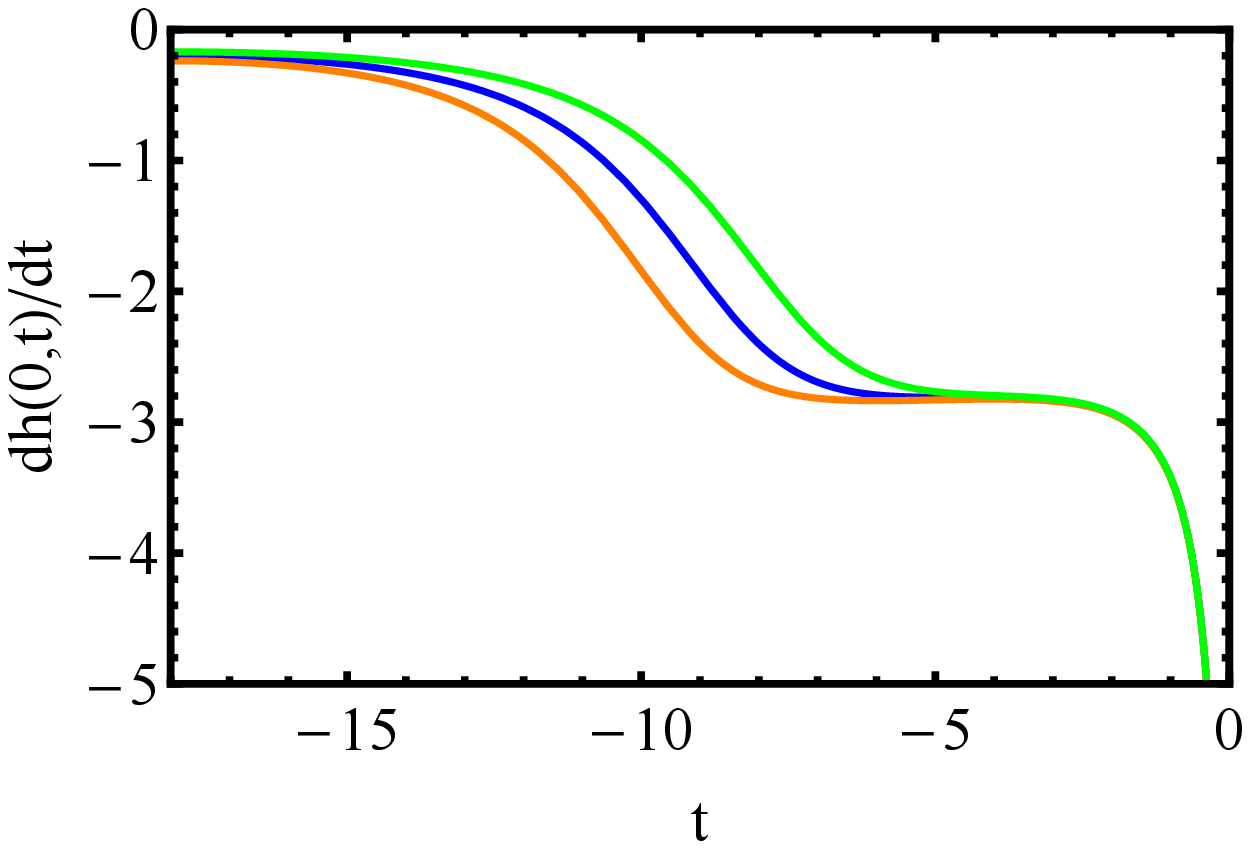}
\caption{$\rho(r=0,t)$ vs. time (left) and $dh(r=0,t)/dt$  vs. time (right) in the regime, corresponding to the $\lambda H>0$ tail of $\mathcal{P} (H)$. The OFM problem was solved numerically for
three different $\Lambda$, corresponding to $H =-29.5$, $-32.5$ and $-35.1$. The plateau regime is well described by the ansatz~(\ref{ansatz1}). The plateau duration increases with $|H|$. For these heights it was sufficient to start the numerical calculations at $t=-18$.}
\label{rhodhdt}
\end{figure}

\begin{figure} [ht]
\includegraphics[width=0.45\textwidth,clip=]{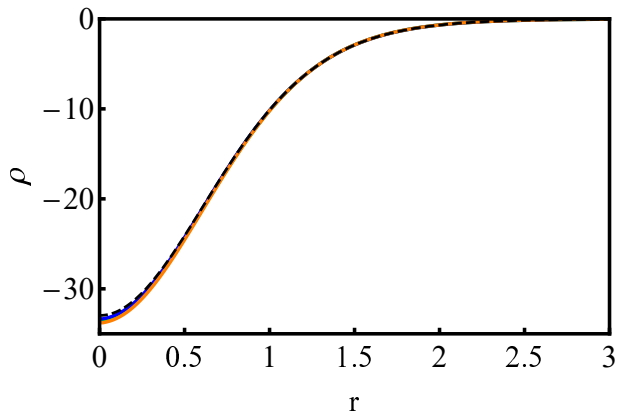}
\caption{Solid lines: Spatial profiles of $\rho(r,t)$, obtained for the same parameters as in Fig. \ref{rhodhdt}. The profiles are shown at $t=-5$ or
$t=-4$, corresponding to the plateau of $\rho(0,t)$ observed in Fig.~\ref{rhodhdt}. Dashed line: $\mathrm{P}\left(r\right)$, obtained from a numerical solution of Eq.~(\ref{H234}) for $c=2.68$, which corresponds to the minimum of the function $f(c)=\dot{s}/c$.}
\label{rhor3}
\end{figure}

Figures~\ref{rhodhdt} and \ref{rhor3} show our numerical results  for sufficiently large negative $H$, or $\Lambda$.\footnote{\label{numerics} We solved the  full OFM problem numerically for the noise correlator~(\ref{correlator}) at $d=3$ by using the Chernykh-Stepanov back-and-forth iteration algorithm \cite{CS} and assuming spherical symmetry. See the Appendix for some details.} Figure~(\ref{rhodhdt}) depicts the numerically found $\rho(r=0,t)$ (left panel) and $dh(r=0,t)/dt$ (right panel) vs. time for three different, and sufficiently large, values of $H<0$. As one can see, the time-dependent solution exhibits three asymptotics: growth in the initial and final stages and a plateau in between. Strikingly, these asymptotics  are identical for different $H$, except that the durations of the plateau are different (they increase with $|H|$), whereas the plateau values coincide.  Crucially, the plateau region corresponds to a \emph{unique} pair of the $\rho$-soliton and $h$-front, described by Eq.~(\ref{ansatz1}). This soliton-front solution has $c\simeq 2.83$. Three solid lines in Fig.~\ref{rhor3}  show the spatial profiles of the solitons, observed at these three different $H$. The collapse of data confirms that this is the same soliton. In the rest of this Section we will show that this unique soliton-front solution is selected because its action is minimum.

Let us consider the soliton-front solutions in some detail. Plugging the ansatz~(\ref{ansatz1}) into Eqs.~(\ref{heqd}) and~(\ref{peqd}), we obtain
\begin{equation}\label{H170}
\nabla^2\mathrm{H}-\frac{1}{2}\left(\nabla \mathrm{H}\right)^2  +\int d\mathbf{x}^{\prime}\, \kappa(\mathbf{x}-\mathbf{x}^{\prime})\, \mathrm{P}(\mathbf{x}^{\prime})+c =0\,,
\quad\quad
\nabla\cdot\left(\nabla \mathrm{P}+\mathrm{P}\nabla \mathrm{H}\right)=0\, .
\end{equation}
It is natural to set $\mathrm{H}(0)=0$. At large distances $\mathrm{P}(\mathbf{x})\to 0$, but  $\mathrm{H}(\mathbf{x})$ can grow indefinitely. The latter feature does not present a problem, as the solution
(\ref{ansatz1}) can be continuously matched with the solution of Eq.~(\ref{Hopf1}) which vanishes at $\mathbf{x}\to \infty$. The resulting composite solution, however,  does not affect the action $s(H)$, so we will not present it here.

Now let us assume that $\mathrm{P}$ is a function of $\mathrm{H}$\footnote{\label{barotropic} This assumption holds automatically for spherically-symmetric solutions which we will focus on shortly.}. Integrating the second equation in~(\ref{H170}), we obtain
$\mathrm{P}(\mathbf{x})=-\mathrm{P}_0\, e^{-\mathrm{H}(\mathbf{x})}$, where $\mathrm{P}_0$ is a positive constant. Plugging this relationship into the first equation in~(\ref{H170}), we obtain
\begin{equation}\label{H170a}
\nabla^2 \mathrm{H} -\frac{1}{2}\left(\nabla\mathrm{H}\right)^2-\mathrm{P}_0\, \int d\mathbf{x}^{\prime}\, \kappa(\mathbf{x}-\mathbf{x}^{\prime}) \, e^{-\mathrm{H}\left(\mathbf{x}^{\prime}\right)}+c =0\, .
\end{equation}
Now we change the variables from  $\mathbf{x}$ and $\mathrm{H}$  to
\begin{equation}\label{changevar}
\boldsymbol{\chi}=\sqrt{\frac{c}{2}} \,\mathbf{x}\quad
\mbox{and} \quad Z(\boldsymbol{\chi})=\sqrt{\frac{P_0}{c}}\,e^{-\frac{1}{2}\,\mathrm{H}(\sqrt{\frac{2}{c}}\,\chi)}\,.
\end{equation}
This brings Eq.~(\ref{H170a}) to the form
\begin{equation}\label{H234}
 \nabla^2 Z(\boldsymbol{\chi})=Z(\boldsymbol{\chi})-Z(\boldsymbol{\chi})\, \int d\boldsymbol{\chi}^{\prime} \, \bar{\kappa}_{c}(\boldsymbol{\chi}-\boldsymbol{\chi}^{\prime}) \, Z^2\left(\boldsymbol{\chi}^{\prime}\right)\, ,
\end{equation}
where
\begin{equation}\label{corrnew}
\bar{\kappa}_{c}(\mathbf{w})=\left(\frac{2}{c}\right)^{d/2} \kappa \left(\sqrt{\frac{2}{c}}\,\mathbf{w}\right)\, ,
\end{equation}
and $\int d\mathbf{w}\, \bar{\kappa}_c(\mathbf{w})=1$.  Equation~(\ref{H234}) is a nonlinear integro-differential equation in partial derivatives.
It should be solved with the boundary condition $Z(\infty)=0$ subject to constraint $Z(\boldsymbol{\chi})>0$.
A solution exists, for any $c>0$, for a single value of $Z(0)=\sqrt{\mathrm{P}_0/c}$. Once the solution is found, one can use it to evaluate
the action accumulation rate~(\ref{sdotdef}), which can be rewritten as
\begin{equation}\label{NW180}
\dot{s}(c)=\frac{1}{2}\int d\mathbf{x}\, d\mathbf{x}^\prime\,
\kappa (\mathbf{x}-\mathbf{x}^\prime) P(\mathbf{x})\,P(\mathbf{x}^\prime)
=\frac{c^2}{2}\left(\frac{2}{c}\right)^{d/2}
\int d\boldsymbol{\chi}\, d\boldsymbol{\chi}^\prime\,\bar{\kappa}_c(\boldsymbol{\chi}-\boldsymbol{\chi^\prime})Z^2(\boldsymbol{\chi})\, Z^2(\boldsymbol{\chi^\prime}) .
\end{equation}

From now on we will assume spherical symmetry, $Z(\boldsymbol{\chi})=Z(\chi)$, where $\chi$ is the (rescaled) radial coordinate. The resulting nonlinear integro-differential equation~(\ref{H234}) still cannot be solved analytically. We made analytical progress in the limits of $c\ll 1$ and $c\gg 1$, and also solved Eq.~(\ref{H234}) numerically.

\subsection{$c\ll 1$: broad solitons}

\subsubsection{Leading order, $2<d<4$}
\label{sss}

For $c\ll 1$, the soliton width is much larger than the correlation length of the noise. Therefore, in the leading order, one can approximate $\bar{\kappa}_c(\mathbf{w})$ by the delta-function, as if the noise were white in space. As a result, (the spherically symmetric version of) Eq.~(\ref{H234}) becomes a nonlinear ODE:
\begin{equation}\label{H240}
\frac{1}{\chi^{d-1}} \frac{d}{d\chi} \left(\chi^{d-1} \frac{dZ}{d\chi}\right) = Z-Z^3 ,
\end{equation}
whereas the $c$-dependence  enters only through the change of variables~(\ref{changevar}).  Equation~(\ref{H240}) should be solved subject to the conditions $Z^{\prime}(0)=0$, $Z(\infty)=0$ and $Z(\chi)>0$, whereas $Z(0)$ is \emph{a priori} unknown.

Equation~(\ref{H240}) has been encountered
in different physical contexts, among them nonlinear optics \cite{Townes}, decay of a false vacuum in theories of a single scalar field \cite{Coleman1977} and calculations of the density of states in disordered media \cite{Yaida}.  This equation is exactly soluble only for $d=1$. The existence of a nontrivial solution vanishing at infinity was proved only for $d<4$ \cite{Coleman}. As we found numerically, for $d\geq 4$ such a solution does not exist. Furthermore, as $d$ (treated as a real positive number) approaches $4$ from below, $Z(0)$ diverges, while the characteristic width of $Z(\chi)$ goes to zero. As a result, at $d<4$ and sufficiently close to $4$, and for larger $d$, the approximation of $\bar{\kappa}_{c}(\mathbf{w})$ by the delta-function breaks down, and one needs
to deal with the integral term in Eq.~(\ref{H234}). The integral term regularizes the solution, so that Eq.~(\ref{H234}) has the required solution at any $d$ and for any $c\geq 0$.  Our analysis in this subsection, however, employs Eq.~(\ref{H240}) and, therefore, is limited to $d<4$.

\begin{figure}[h]
\includegraphics[width=0.43\textwidth,clip=]{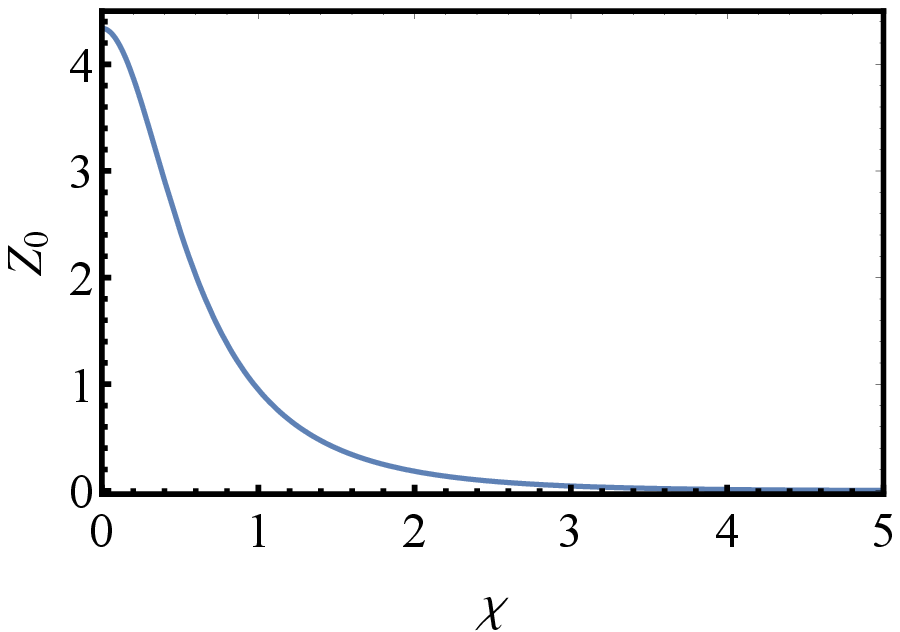}
\includegraphics[width=0.45\textwidth,clip=]{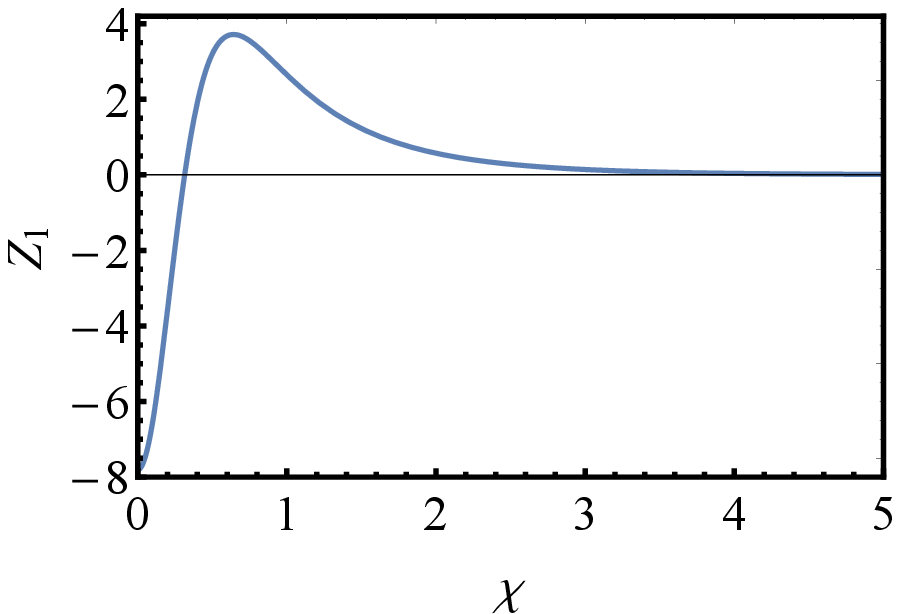}
\caption{Numerical solutions of Eq.~(\ref{H240}) (left panel) and (\ref{frstordrc}) (right panel) for $d=3$. Here $Z(\chi)=Z_0(\chi)+c Z_1(\chi) +\dots$, where $c\ll 1$.}
\label{3Z}
\end{figure}

Equation~(\ref{H240}) can be easily solved numerically by the shooting method, using $Z(0)$ as the shooting parameter.
The left panel of Fig. \ref{3Z} depicts the numerically found $Z=Z_0(\chi)$ for $d=3$. Here we obtained
$Z_0(0)\simeq 4.3374.$

Now we can evaluate, in the leading order in $c\ll 1$, the action accumulation rate $\dot{s}(c)$. Using Eq.~(\ref{NW180}) with
$\bar{\kappa}_c(\boldsymbol{\chi}-\boldsymbol{\chi}^{\prime})$ replaced by the delta-function, we obtain
\begin{equation}\label{sdotwide2}
\dot{s}(c) = k_d c^{\frac{4-d}{2}} \,,\quad \quad c\ll 1\,, \quad 2<d<4\,,
\end{equation}
where
\begin{equation}\label{kd}
k_d = \frac{(2\pi)^{d/2}}{\Gamma(d/2)} \int_0^{\infty}d\chi\, \chi^{d-1} \,Z_0^4(\chi) .
\end{equation}
Here we used the expression $\sigma_{d-1} = 2\pi^{d/2}/\Gamma(d/2)$ for the surface area of the sphere of unit radius in $\mathbb{R}^d$.
Once the numerical solution $Z_d(\chi)$ is known, the integral in Eq.~(\ref{kd}) can be evaluated numerically.  For $d=3$ we obtain $k_3\simeq 106.9$.

Suppose that the $\rho$-soliton acts for time $\tau$, and the traveling $h$-front~(\ref{ansatz1}) reaches the specified height $H$, we obtain $\tau \simeq |H|/c$. As a result, for $d=3$,
\begin{equation}\label{swide}
s(H)\simeq \dot{s}(c) \,\tau = \frac{k_3}{\sqrt{c}} \,|H|\,,\quad c\ll 1\,.
\end{equation}

\subsubsection{Subleading order, $2<d<4$}

Now let us return to Eq.~(\ref{H234}) and calculate the correction $O(c)$ to $Z_0(\chi)$ by Taylor-expanding $Z^2(\boldsymbol{\chi}^{\prime})$ under the integral in the vicinity of $\boldsymbol{\chi}^{\prime}=\boldsymbol{\chi}$. The zeroth-order term
yields, after the integration, the term $-Z^3$ as in Eq.~(\ref{H240}). The first-order terms
do not contribute to the integral. The first non-vanishing correction, $O(c)$, comes
from the symmetric second-order terms of the Taylor series, and we obtain
\begin{equation}\label{corr0}
\int d\boldsymbol{\chi}^{\prime} \, \bar{\kappa}_{c}(\boldsymbol{\chi}-\boldsymbol{\chi}^{\prime}) \, Z^2\left(\boldsymbol{\chi}^{\prime}\right) \simeq Z^2(\boldsymbol{\chi})+\frac{c \delta^2}{4d} \nabla^2 Z^2(\boldsymbol{\chi}),
\end{equation}
where
\begin{equation}\label{p010}
\delta^2=\int \mathbf{w}^2 \kappa\left(\mathbf{w}\right)\, d\mathbf{w} =O(1)
\end{equation}
depends on the particular form of the noise correlator.  [Note that it is $\kappa$, not $\bar{\kappa}_c$, which enters the definition~(\ref{p010}.)] As a result, Eq.~(\ref{H234}) takes the form
\begin{equation}\label{corr1}
\nabla^2 Z=Z-Z^3-\frac{c\delta^2}{4d}\, Z\, \nabla^2 Z^2\,.
\end{equation}
For spherically-symmetric solutions at $d=3$, and for the correlator (\ref{correlator}), for which $\delta^2=d/2$, Eq.~(\ref{corr1}) becomes
\begin{equation}\label{p020}
\frac{d}{d\chi}\left(\chi^2 \frac{dZ}{d\chi}\right) = \chi^2 (Z-Z^3) - \frac{c}{4}  Z \frac{d}{d\chi}\left(\chi^2 Z \frac{dZ}{d\chi}\right).
\end{equation}
Now we can set $Z(\chi)=Z_0(\chi)+ c Z_1(\chi)$ and treat the last term in Eq.~(\ref{p020}) perturbatively. In the leading order we reproduce Eq.~(\ref{H240}) for $Z_0$. The subleading order yields a forced linear equation for $Z_1$:
\begin{equation}\label{frstordrc}
\frac{d}{d\chi}\left(\chi^{2}\frac{dZ_1}{d\chi}\right)+\chi^{2}(3Z_{0}^{2}-1)Z_1=-\frac{Z_{0}}{4}\frac{d}{d\chi}\left(\chi^{2}Z_0 \frac{dZ_0}{d\chi}\right).
\end{equation}
This equation should be solved with the boundary conditions $Z_1^{\prime}(0)=Z_1(\infty)=0$. The solution, obtained by shooting, is shown in the right panel of Fig. \ref{3Z}.  Here $Z_1(0)\simeq -7.805$.

Now we evaluate, in the subleading order,  $\dot{s}$ from Eq.~(\ref{NW180}). The integration over $\boldsymbol{\chi}^{\prime}$ was already performed, see Eq.~(\ref{corr0}). For $d=3$ and the correlator (\ref{correlator}) we obtain
\begin{equation}\label{sdotcorr1}
\dot{s}(c) =\sqrt{2c} \int d \boldsymbol{\chi} Z^4(\boldsymbol{\chi})+\sqrt{2c}\,\frac{3c}{8} \int d \boldsymbol{\chi} Z^3(\boldsymbol{\chi})\nabla^2 Z^2 (\boldsymbol{\chi}).
\end{equation}
Now we set $Z(\chi)=Z_0(\chi)+ c Z_1(\chi)$ and obtain, in the leading and subleading orders,
\begin{equation}\label{sdotcorr2}
\dot{s}(c) = \sqrt{2c} \int d \boldsymbol{\chi} Z_0^4(\boldsymbol{\chi})+\sqrt{2c} \cdot 3c \int d \boldsymbol{\chi} Z_0^3(\boldsymbol{\chi})Z_1(\boldsymbol{\chi})+\sqrt{2c}\cdot\frac{3c}{8}
\int d \boldsymbol{\chi} Z_0^3(\boldsymbol{\chi})\nabla^2 Z_0^2 (\boldsymbol{\chi}).
\end{equation}
The first term  brings us back to the leading-order results~(\ref{sdotwide2}) and (\ref{kd}) (for $d=3$), as to be expected. The second and third terms give the subleading contribution, $137.7 \,c^{3/2}$. Collecting the leading and subleading terms, we obtain
\begin{equation}\label{swide10}
s(H)\simeq \dot{s}(c) \,\tau \simeq \left(\frac{106.9}{\sqrt{c}} +137.7 \sqrt{c} +\dots  \right) \,|H| \,,\quad c\ll 1\,.
\end{equation}

\subsection{$c\gg 1$: narrow solitons}
For $c\gg 1$ the soliton width is much smaller than the correlation length of the noise.
In the language of Eq.~(\ref{H234}), $\bar{\kappa}_{c}$ is almost constant in the region where $Z(\mathbf{\chi})$ is not exponentially small. Therefore,  Eq.~(\ref{H234}) simplifies:
\begin{equation}\label{NW005}
 \nabla^2 Z(\chi)= \left[1-A\, \bar{\kappa}_{c}(\chi) \right]\, Z(\chi)\, ,
\end{equation}
where
\begin{equation}\label{NW034}
A=\int d\boldsymbol{\chi}Z^2(\chi)\, .
\end{equation}
Furthermore, in this region it suffices to expand $\kappa(w)$ to second order in $w$:
\begin{equation}\label{NW010}
\kappa(w)\simeq
\kappa(0)-\frac{1}{2}\kappa^{\prime\prime}(0)\, w^2\, .
\end{equation}
In view of Eq.~(\ref{corrnew}), we have
\begin{equation}\label{NW020}
\bar{\kappa}_c(w)\simeq\left(\frac{2}{c}\right)^{d/2}\,
\left(\kappa(0)-\left|\kappa^{\prime\prime}(0)\right|\, \frac{w^2}{c}\right)\,,
\end{equation}
which we plug into Eq.~(\ref{H234}) and obtain
\begin{equation}\label{NW030}
 \nabla^2  Z(\chi)+
  \left[\left(\frac{2}{c}\right)^{d/2}\kappa(0)A -1 -
 \frac{1}{2}\left(\frac{2}{c}\right)^{\frac{d+2}{2}}\left|\kappa^{\prime\prime}(0)\right|A\, \chi^2\right]\, Z(\chi)=0\, .
\end{equation}
For a fixed $A$ this is a Schroedinger equation for an isotropic $d$-dimensional harmonic oscillator,
\begin{equation}\label{Schr}
\nabla^2 Z +\frac{2m}{\hbar^2}\left(E-\frac{m \omega^2 \chi^2}{2}\right) Z=0\,,
\end{equation}
where we can set
\begin{equation}\label{ident}
\hbar=1,\quad m=1/2, \quad E=\left(\frac{2}{c}\right)^{d/2}\kappa(0)A -1\quad\mbox{and}\quad \omega^2 = 2 \left(\frac{2}{c}\right)^{\frac{d+2}{2}} |\kappa^{\prime\prime}(0)| A\,.
\end{equation}
As $Z(\chi)$ must be everywhere positive, we should find the ground-state wave function. The ground state energy is $E=(d/2)\,\hbar \omega$. This yields an algebraic equation for $A$, which solution, at $c\gg 1$, is
\begin{equation}\label{NW060}
  \left(\frac{2}{c}\right)^{d/2}A=\frac{1}{\kappa(0)}\left(1+
 d\sqrt{\frac{\left|\kappa^{\prime\prime}(0)\right|}
  {\kappa(0)c}}+\dots\right)\,.
\end{equation}
As a result,
\begin{equation}\label{NW070}
\omega\simeq
2\sqrt{\frac{\left|\kappa^{\prime\prime}(0)\right|}
  {\kappa(0)c}}
\end{equation}
and
\begin{equation}\label{NW090}
Z(\chi) =B\, \exp\left(-\frac{\chi^2}{2}\sqrt{\frac{\left|\kappa^{\prime\prime}(0)\right|}
  {\kappa(0)c}}\right)\,.
\end{equation}
The constant $B$ can found
from Eq.~(\ref{NW034})\footnote{\label{outer} The solution~(\ref{NW090}) is valid for $\chi\ll c^{1/2}$, where the expansion~(\ref{NW020}) of $\bar{\kappa}_c(w)$ still holds. It can be complemented
by an evanescent WKB solution of Eq.~(\ref{NW005}) which is valid at distances $\chi\gg c^{1/4}$. The two solutions can be matched in their joint region $c^{1/4}\ll \chi \ll c^{1/2}$.}
but, in fact, there is no need in the explicit form of $Z(\chi)$ for the purpose of calculating the action in the leading order of $c\gg 1$.
Indeed, let us evaluate the double integral in Eq.~(\ref{NW180}) for $\dot{s}$. As one can see from Eq.~(\ref{NW090}), $Z(\chi)$ is localized in a region of the size $\sim c^{1/4}$,
whereas the characteristic length scale of $\bar{\kappa}_c(w)$ from Eq.~(\ref{NW020}) is $\sim c^{1/2} \gg c^{1/4}$. Therefore, we can replace in Eq.~(\ref{NW180}) $\bar{\kappa}_c(\boldsymbol{\chi}-\boldsymbol{\chi}^{\prime}$ by $\bar{\kappa}_c(0)$ and obtain
\begin{equation}\label{NW170a}
\int d\boldsymbol{\chi}\, d\boldsymbol{\chi}^\prime\, Z^2(\chi)\bar{\kappa}_c(\boldsymbol{\chi}-
\boldsymbol{\chi}^\prime) Z^2(\chi^\prime)
\simeq\bar{\kappa}_c(0) \int d\boldsymbol{\chi}\, Z^2(\chi) \,\int d\boldsymbol{\chi^{\prime}}\, Z^2(\chi^{\prime})
= \bar{\kappa}_c(0) A^2
\simeq \frac{1}{\kappa(0)} \left(\frac{c}{2}\right)^{d/2} \,,
\end{equation}
leading to
\begin{equation}\label{sdotnarrow}
\dot{s}(c) = \frac{c^2}{2 \kappa(0)}\,, \quad \quad c\gg 1\,.
\end{equation}
Suppose that the soliton acts for time $\tau$, during which the traveling $h$-front reaches the height $H$. Therefore, $\tau\simeq |H|/c$ and
\begin{equation}\label{snarrow}
s(H)\simeq \dot{s}(c) \,\tau = \frac{|H| c}{2 \kappa(0)}\,.
\end{equation}

\subsection{Selection of $c$}

\begin{figure} [ht]
\includegraphics[width=0.45\textwidth,clip=]{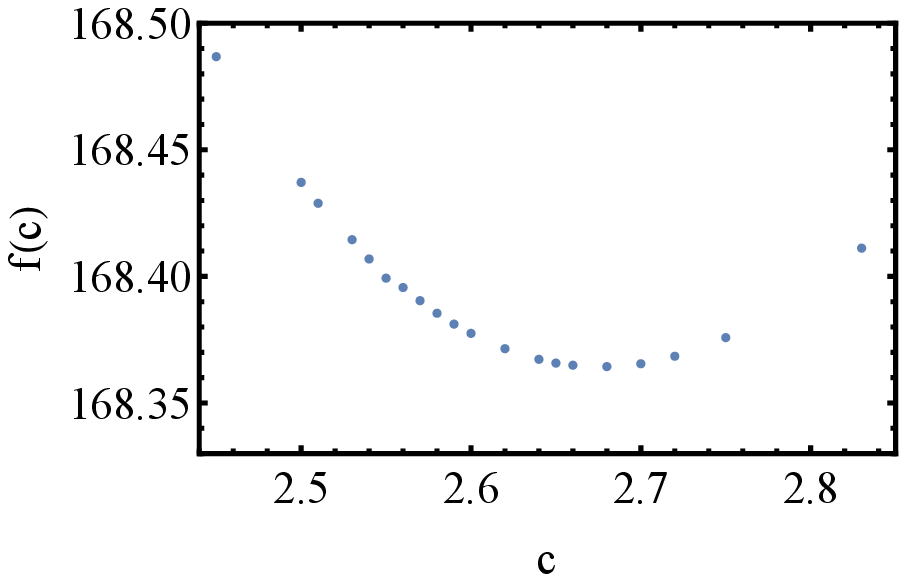}
\caption{The function $f(c)=\dot{s}/c$ vs. $c$ in the region of its minimum for $d=3$. $f(c)$ was calculated by numerically solving  Eq.~(\ref{H234}) for the correlator~(\ref{correlator}) and numerically evaluating the integral in Eq.~(\ref{NW180}).}
\label{svsc}
\end{figure}

\begin{figure} [ht]
\includegraphics[width=0.45\textwidth,clip=]{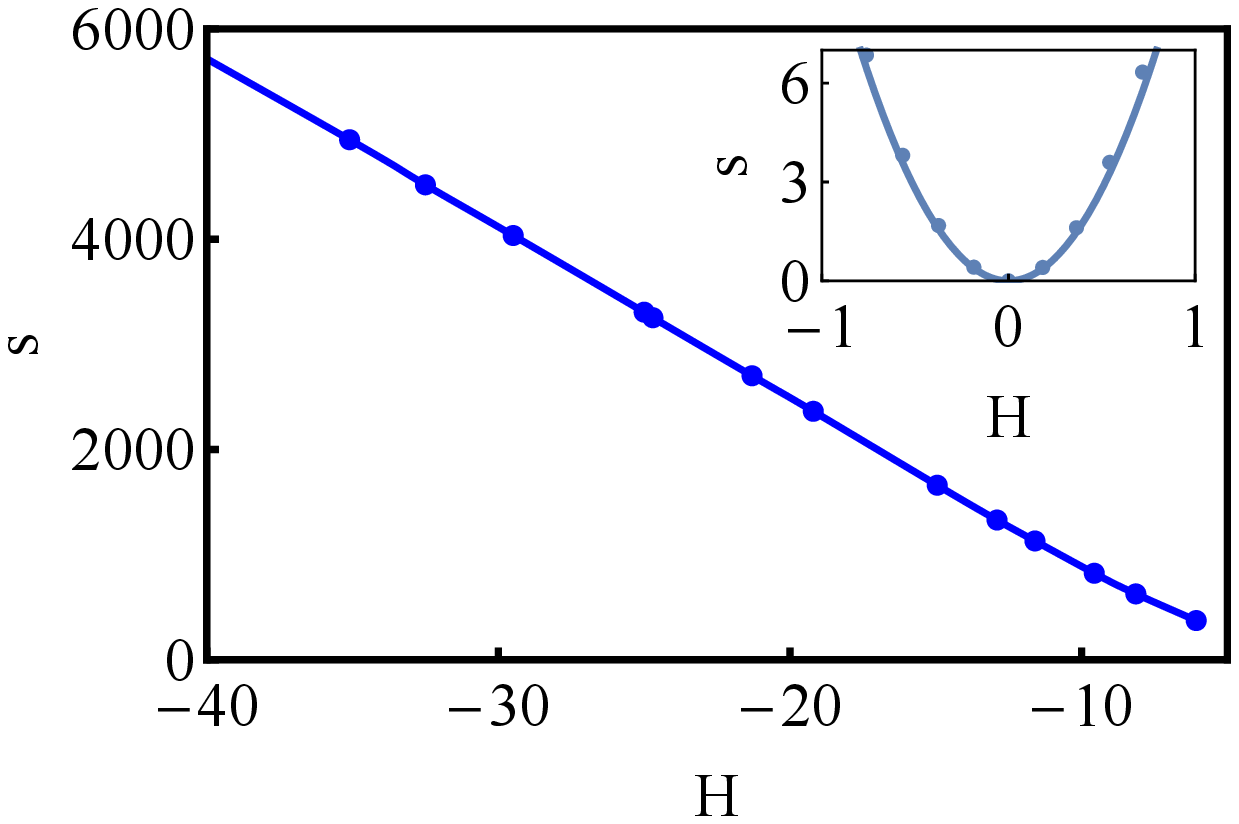}
\includegraphics[width=0.47\textwidth,clip=]{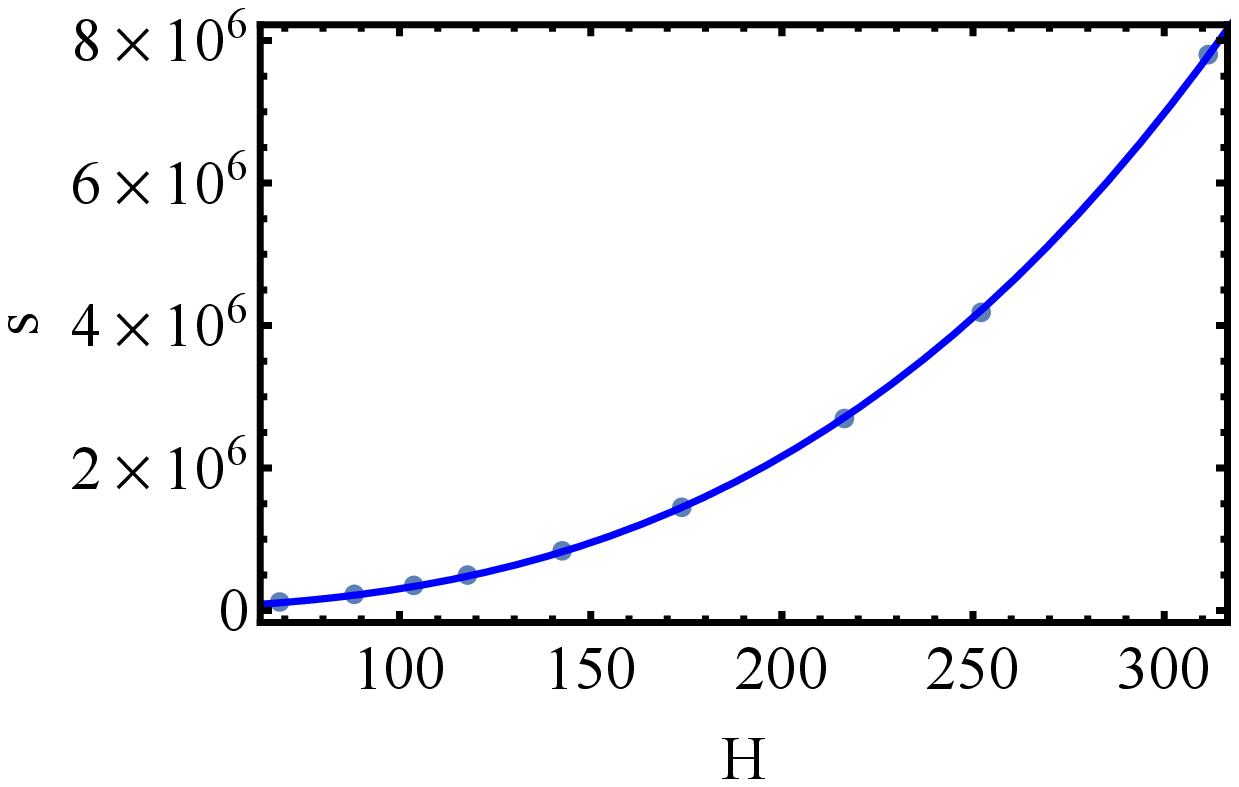}

\caption{The rescaled action  $s=s(H)$, see Eq.~(\ref{sd}), for the correlator~(\ref{correlator}) at $d=3$. Left panel: $H<0$. The slope of the straight part of the line is close to $163$, which corresponds to Eq.~(\ref{slowtail}) with $C_s=163$.  Right panel: $H>0$. Shown are numerical results (symbols) and the fit $s(H)\simeq  0.23 H^3+7.12 H^2$ which agrees with  Eq.~(\ref{fasttail}) with $C_f\simeq 0.23$. Inset of the left panel: $s(H)$ for small $|H|$; here the numerical results (symbols) agree with predictions from the linear theory, Eqs.~(\ref{variance}) and~(\ref{C0example}) (solid line).}
\label{svsH}
\end{figure}

At arbitrary $c$, the action $s(H)$ can be written, for very large $\lambda H>0$, as $s(H)=|H|\,f(c)$.  The $c\ll 1$ and $c\gg 1$ asymptotics of $f(c)$, as given by Eqs.~(\ref{swide10}) and (\ref{snarrow}), are the following (for $d=3$):
\begin{equation}
\label{falpha}
f\left(c\right)=\frac{\dot{s}(c)}{c}\simeq \begin{cases}
\frac{106.9}{\sqrt{c}} +137.7 \sqrt{c} +\dots ,& c\ll 1,\\
\frac{\pi^{3/2} c}{2} + \dots, & c\gg 1\,.
\end{cases}
\end{equation}
The selected value of $c$ is determined by the minimum of $f(c)$ as a function of $c$. The minimum is located in the intermediate region $c=O(1)$, where neither of the asymptotics (\ref{falpha}) is valid. Therefore, we solved Eq.~(\ref{H234})  numerically, using an algorithm briefly described in the Appendix. Then we evaluated the action accumulation rate $\dot{s}$ from Eq.~(\ref{NW180}), and the function $f(c)=\dot{s}/c$, for different $c$. At very small and very large $c$ our numerical results for $f(c)$ agree well with the asymptotics (\ref{falpha}). A numerical graph of $f(c)$ for intermediate $c$ is shown in Fig.~\ref{svsc}. The minimum
is observed at $c\simeq 2.68$, in a fair agreement with $c\simeq 2.8$, observed in the plateau region of Fig.~\ref{rhodhdt}. The left panel of Fig.~~\ref{svsH} shows the action~(\ref{sd}) as a function of $H<0$, obtained by solving numerically the full time-dependent OFM equations. As one can see, at large negative $H$ the action $s$ is proportional to $|H|$. The proportionality coefficient [which yields the coefficient $C_s$ in Eq.~(\ref{slowtail})] is close to $163$. This value agrees fairy well with the theoretical minimum value $f(c=2.68)=168.36$, obtained from the numerical solution of Eq.~(\ref{H234}).

\begin{figure} [ht]
\includegraphics[width=0.45\textwidth,clip=]{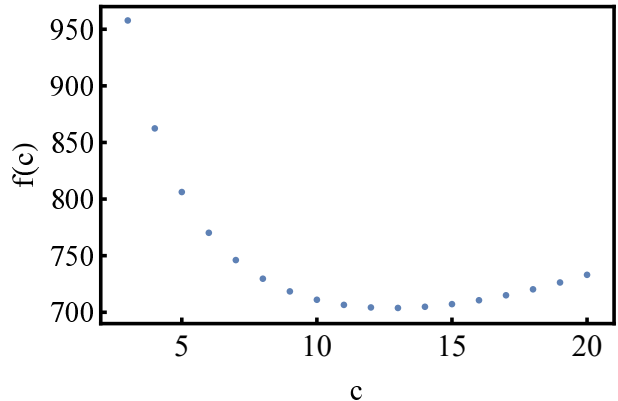}
\caption{The function $f(c)=\dot{s}/c$ vs. $c$ in the region of its minimum for $d=5$. $f(c)$ was calculated by numerically solving  Eq.~(\ref{H234}) for the correlator~(\ref{correlator}) and numerically evaluating the integral in Eq.~(\ref{NW180}).}
\label{d=5fig}
\end{figure}

For $d\geq 4$ a different asymptotic expansion at $c\ll 1$ is required, as explained in subsubsection \ref{sss}. Still, our numerics shows that, similarly to the case of $2<d<4$, the minimum action rate is achieved at an intermediate value of $c$. This is illustrated by Fig. \ref{d=5fig}, obtained for $d=5$.

To summarize this Section, the action minimization with respect to $c$ selects  $c=O(1)$ uniquely and brings us to the announced result~(\ref{slowtail}) for the slower-decaying tail of $\mathcal{P}(H)$.
The characteristic length scale of the optimal realization of noise $\rho(\mathbf{x})$  is of the order of unity or, in the original variables, of the order of the correlation length of the noise $\ell$ \cite{KK2008}. The characteristic formation time of the distribution tail ~(\ref{slowtail})
scales as the diffusion time over the correlation length and grows linearly with $|H|$.

\section{The $\lambda H <0$ tail}
\label{positive}

Similarly to the case of $d=1$ \cite{KK2007,KK2009,MKV}, the $\lambda H <0$ tail is dominated by the nonlinearity to such an extent that, in the leading order, the tail is independent of $\nu$. Here the optimal realization of the noise field, $\rho(\mathbf{x},t)$, for most of the time, is large-scale, and one can drop the diffusion terms in the OFM equations~(\ref{heqd}) and (\ref{peqd}). Furthermore, during most of the dynamics the characteristic length scale of $h(\mathbf{x},t)$ and $\rho(\mathbf{x},t)$ is much larger than the noise correlation length. Here the function $\kappa(\mathbf{x}-\mathbf{x}^{\prime})$ under the integral in Eq.~(\ref{pi}) for $\pi(\mathbf{x},t)$ can be approximated by the delta-function, and one arrives at $\pi(\mathbf{x},t)\simeq \rho(\mathbf{x},t)$, as if the noise were white in space. Taking the gradient of  Eq.~(\ref{heqd}), we arrive at
equations which describe an inviscid hydrodynamic flow of an effective gas with negative pressure:
\begin{eqnarray}
\partial_t\rho+\nabla \cdot (\rho \mathbf{V})&=&0\, ,
\label{n0100}\\
\partial_t \mathbf{V}+(\mathbf{V}\cdot \nabla) \mathbf{V}&=&\nabla \rho\, ,
\label{n0200}
\end{eqnarray}
where $\rho(\mathbf{x},t)$ can be interpreted as the gas density, and the height gradient field
$\mathbf{V}(\mathbf{x},t)=\nabla h(\mathbf{x},t)$ as the gas velocity. Equations~(\ref{n0100}) and (\ref{n0200}) appear in many contexts, where they provide a large-scale description of a plethora of hydrodynamic instabilities \cite{TZ}. In all these problems, however, one deals with an initial-value problem, whereas here we have to deal with a boundary-value problem in time. At $t=0$ we have $\mathbf{V}(\mathbf{x},0)=0$ whereas, in view of Eqs.~(\ref{ptimezero}) and (\ref{conservation}), the solution must describe collapse of a gas cloud of mass $\Lambda$ into the origin. Close to the collapse time $t=0$,  the size of the gas cloud (which we will determine shortly) becomes comparable with the correlation length of the noise, and the term $\nabla \rho(\mathbf{x},t)$  in Eq.~(\ref{n0200}) should be replaced by the nonlocal term $\nabla \pi(\mathbf{x},t)$. But let us first solve Eqs.~(\ref{n0100}) and (\ref{n0200}) as they are. Assuming spherical symmetry,
we rewrite these equations as
\begin{eqnarray}
\partial_t\rho+\frac{1}{r^{d-1}}\partial_r \left(r^{d-1} \rho V\right)&=&0\, ,
\label{n010}\\
\partial_t V+V\partial_r V&=&\partial_r\rho\, ,
\label{n020}
\end{eqnarray}
These equations and the conditions (\ref{timeminusinfinity}) and (\ref{ptimezero}) have no intrinsic length or time scale, so the solution
must be self-similar \cite{Barenblatt}. With account of the conservation law (\ref{conservation}), the similarity exponents can be determined immediately:
\begin{equation}\label{n420}
\rho(r,t)=(-t)^{-\frac{2d}{d+2}}\, R(\xi) ,\quad V(r,t)=(-t)^{-\frac{d}{d+2}}\, W(\xi) ,\quad
\mbox{where}\quad \xi=r\, (-t)^{-\frac{2}{d+2}}\, ,
\end{equation}
and $R(\xi)\geq 0$. Plugging this ansatz into Eqs.~(\ref{n010}) and (\ref{n020}), we arrive at two coupled ODEs for $R(\xi)$ and $W(\xi)$:
\begin{eqnarray}
\left(W+\frac{2 \xi}{d+2}\right) R^\prime +RW^\prime&=&-\frac{2d}{d+2}R-\frac{d-1}{\xi}RW\, ,
\label{n430}\\
R^\prime-\left(W+\frac{2 \xi}{d+2}\right) W^\prime&=&\frac{d}{d+2}W\, .
\label{n440}
\end{eqnarray}
These equations can be treated as linear algebraic equations for the derivatives $R^{\prime}$ and $W^{\prime}$. The derivatives can be determined uniquely in terms of $R(\xi), W(\xi)$ and $\xi$, except in a special point
$$
R=0, \quad W=-\frac{2 \xi}{d+2},
$$
where the determinant of Eqs.~(\ref{n430}) and (\ref{n440}),
\begin{equation}\label{n444}
\text{det}=-\left(\frac{2\xi}{d+2}+W\right)^2-R\,,
\end{equation}
vanishes. Solving the resulting equations for $R^{\prime}$ and $W^{\prime}$, we obtain
two different solutions for $R(\xi)$ and $W(\xi)$: a solution where $R(\xi)$ does not vanish identically, and a solution where it does. The former solution has a compact support, and it is very simple:
\begin{equation}\label{n480}
R(\xi) =\frac{d}{(d+2)^2} (\ell_0^2\,- \xi^2)\, ,\quad W(\xi)=-\frac{2 \xi}{d+2}\,,
\quad \mbox{at} \quad 0\leq \xi\leq \ell_0 \,.
\end{equation}
where $\ell_0>0$ is a temporary parameter that can be expressed through $\Lambda$ and ultimately through $H$. In the language of hydrodynamic analogy, the solution (\ref{n420}) and (\ref{n480}) describes a uniform-strain flow. The radius of the imploding ``gas cloud" decreases with time as
\begin{equation}\label{cloudradius}
r_0(t)=\ell_0 (-t)^{\frac{2}{d+2}}\,.
\end{equation}
It is finite for $-\infty<t<0$, and shrinks to zero at $t=0$. In the particular case $d=3$ the similarity solution (\ref{n420}) and (\ref{n480}) was obtained in Ref. \cite{T1988} in the context of a simplified model of gravitational collapse.

The second similarity solution has $R(\xi)\equiv 0$ and therefore $\rho(r,t)\equiv 0$, so it is deterministic. This flow solves the Hopf equation
$\partial_tV+V\partial_r V = 0$ or, in terms of the similarity variable $\xi$,  the equation
\begin{equation}\label{n460}
WW^\prime+\frac{2 \xi}{d+2}W^\prime+\frac{d}{d+2}W=0\,.
\end{equation}
This equation can be also obtained from Eqs.~(\ref{n430}) and (\ref{n440}) by putting there $R(\xi)=0$. The solution of Eq.~(\ref{n460}) exists at $\xi>\ell_0$, and can be matched continuously with the ``pressure-driven"
solution for $W$ from Eq.~(\ref{n480}). The Hopf solution can be obtained analytically, but we will
not present it here, as it is unnecessary for the purpose of calculation of the action and of the interface height at the origin. We will only notice that the matching point of the internal and external solutions, $\xi=\ell_0$, is exactly the point where the determinant~(\ref{n444}) vanishes\footnote{There are also solutions to Eq.~(\ref{n430}) and (\ref{n440}), where $R(\xi)>0$ at all $\xi$, and the Hopf flow is absent. For these solutions $R(\xi)$ behaves as $\text{const}\,\xi^{-d}$ at $\xi \to \infty$. As a result,  $\int \rho(\mathbf{x},t) \, d\mathbf{x}$ diverges, so these solutions should be ruled out.}.

Now let us see what happens if we attempt to use the similarity solution (\ref{n420}) and (\ref{n480}) for $\rho(r,t)$ all the way to $t=0$ in order to calculate the action $s$ from Eq.~(\ref{sd}), or rather from its simpler
local version, obtained by setting $\pi(\mathbf{x},t)\simeq \rho(\mathbf{x},t)$:
\begin{equation}\label{sdzerocorr}
s\simeq \frac{1}{2}\int_{-\infty}^0 dt \int d\mathbf{x} \, \rho^2(\mathbf{x},t) =
\frac{\sigma_{d-1}}{2}\int_{-\infty}^0 dt \int_0^{r_0(t)} dr\,r^{d-1} \, \rho^2(r,t)\,.
\end{equation}
The change of variable $r \to \xi$ yields
\begin{equation}\label{sdzerocorr1}
s= \frac{\sigma_{d-1}}{2}\int_{-\infty}^0 dt\, (-t)^{-\frac{2d}{d+2}}\int_0^{\ell_0} d\xi\,\xi^{d-1} \, R^2(\xi)\,.
\end{equation}
The integral over $\xi$ is well behaved, but the integral over time diverges at $d\geq 2$ at the upper limit $t=0$.

The interface height at the origin at $t=0$ also diverges in this case. In order to show it, we first notice that $r=0$ is the point of maximum of $h(r,t)$. Using Eq.~(\ref{heqd}) (with $\pi$ replaced by $\rho$) in the inviscid limit, we obtain
\begin{equation}\label{dhdt0}
\partial_t h(0,t) = \pi(0,t) \simeq \rho(0,t) = \ell_0^2 (-t)^{-\frac{2d}{d+2}} .
\end{equation}
To obtain $h(0,0)$ we should integrate this expression over time from $-\infty$ to $0$,  but at $d\geq 2$ this integral diverges at the upper limit $t=0$, in exactly the same way as in Eq.~(\ref{sdzerocorr1}). The divergences of $s$ and $h(0,0)$ as functions of $\ell_0$ or $\Lambda$ reflect the fact
that $\mathcal{P}(H)$ is ill-defined when the noise is delta-correlated in space. In the context of short-time (non-stationary) statistics of the interface height of the KPZ equation these divergences were recognized earlier \cite{KK2009}.

For a finite spatial correlation length of the noise $s(H)$ is well defined. An accurate analytic theory would require solving Eqs.~(\ref{heqd}) and (\ref{peqd}) without the
simplifying assumption $\pi(\mathbf{x},t) \simeq \rho(\mathbf{x},t)$ which breaks down close to
$t=0$.  In the absence of such a theory, we can obtain fairly satisfactory results by introducing a cutoff
in the similarity solution (\ref{n420}) and (\ref{n480}) at time $t_c<0$ such that the radius of the ``gas cloud" $r_0(t_c)$ becomes comparable with the noise correlation length [which, in the rescaled units, is $O(1)$]. By virtue of Eq.~(\ref{cloudradius}), one has $-t_c \sim \ell_0^{-(d+2)/2}$. Performing integrations over time from $-\infty$ to $t_c$,
we obtain
\begin{equation}\label{sandH}
s \sim \ell_0^{\frac{3(d+2)}{2}}\quad \mbox{and} \quad H \sim \ell_0^{\frac{d+2}{2}}\,,
\end{equation}
up to numerical pre-factors of order unity, which depend on $d$ and on the exact form of the noise correlator. Eliminating $\ell_0$ from Eq.~(\ref{sandH}), we obtain $s = C_f H^3$, with  $C_f=O(1)$. In view of the scaling relation (\ref{ansatz}) this leads to the announced result~(\ref{fasttail}) for the faster-decaying tail of $\mathcal{P}(H)$. As the rest of $\mathcal{P}(H)$, this tail is mostly contributed to by the noise with the length scale of order $\ell$.

The pre-factors $O(1)$ can be determined numerically, by using the Chernykh-Stepanov back-and-forth iteration algorithm \cite{CS}. A numerical solution is necessarily finite-time. Because of the power-law behavior of the similarity solution (\ref{n420}) as a function of time,  the rescaled action $s$ and height $h(\mathbf{x}=0,t=0)=H$ converge to their steady-state values only \emph{algebraically} in time. For $d=3$ this convergence is quite slow,  as $t^{-1/5}$.  In order to avoid prohibitively long simulations, we determined the steady-state values of $s$ and $h(0,0)$ for each $\Lambda$ by extrapolating  our finite-time numerical results to $t=\infty$. The right panel of Fig.~\ref{svsH} shows the resulting dependence $s=s(H)$ for sufficiently large positive $H$, for the noise correlator~(\ref{correlator}) at $d=3$. The numerics confirm the leading-order cubic behavior of $s(H)$ and yield $C_f \simeq 0.23$.

\begin{figure} [ht]
\includegraphics[width=0.45\textwidth,clip=]{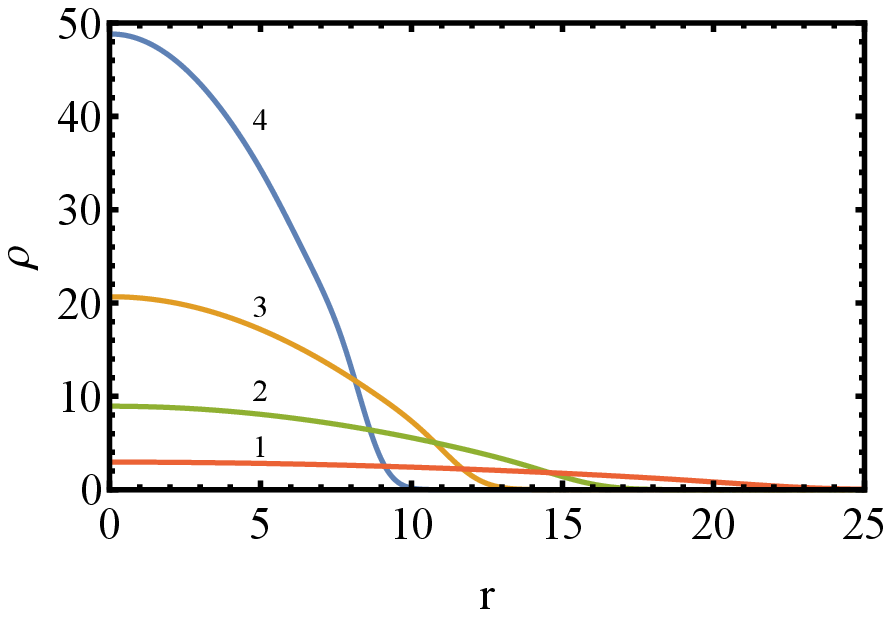}
\includegraphics[width=0.45\textwidth,clip=]{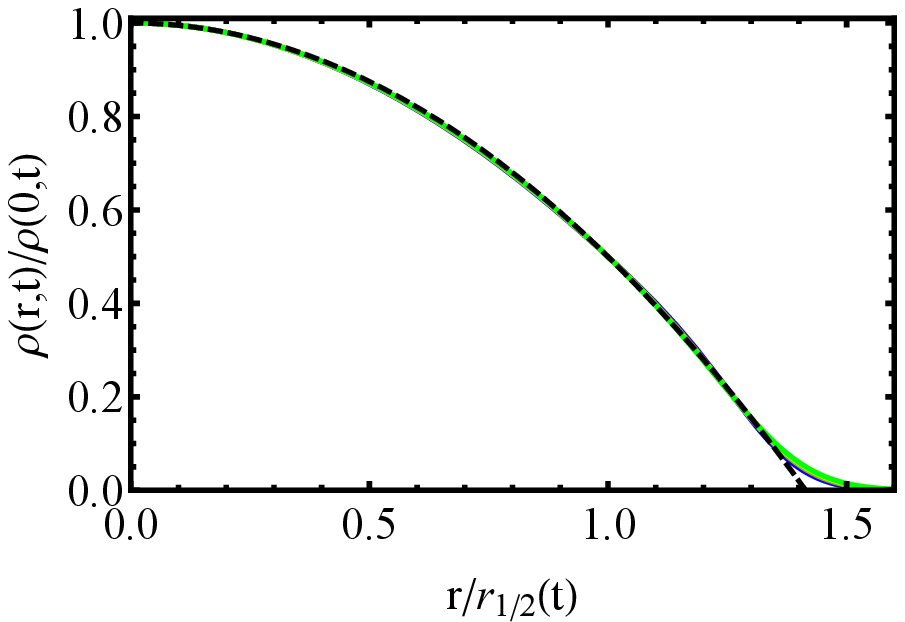}
\caption{Left panel: Numerically computed $\rho(r,t)$ as a function of $r$ for times $t=-5$ (1), $-2$ (2), $-1$ (3) and $-0.5$ (4) in the regime corresponding to the $\lambda H<0$ tail.  Right panel: The same four profiles in rescaled coordinates:  $\rho(r,t)/\rho(0,t)$ versus $r/r_{1/2}(t)$. Dashed line: theoretical prediction (\ref{n480}). In this example $H\simeq 287$, $s\simeq 7.25 \times 10^6$ and $\ell_0\simeq 12.7$.}
\label{HDsim1}
\end{figure}

\begin{figure} [ht]
\includegraphics[width=0.45\textwidth,clip=]{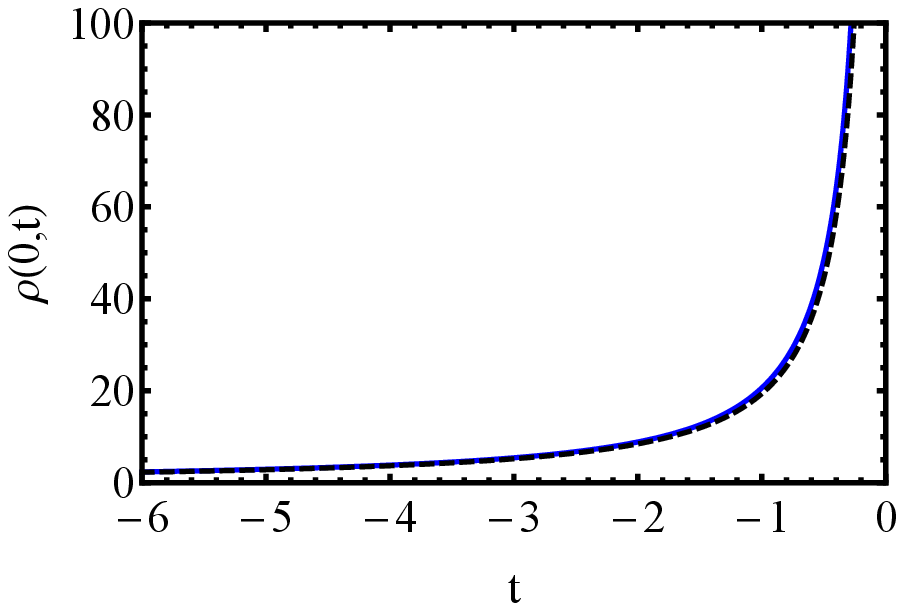}
\includegraphics[width=0.45\textwidth,clip=]{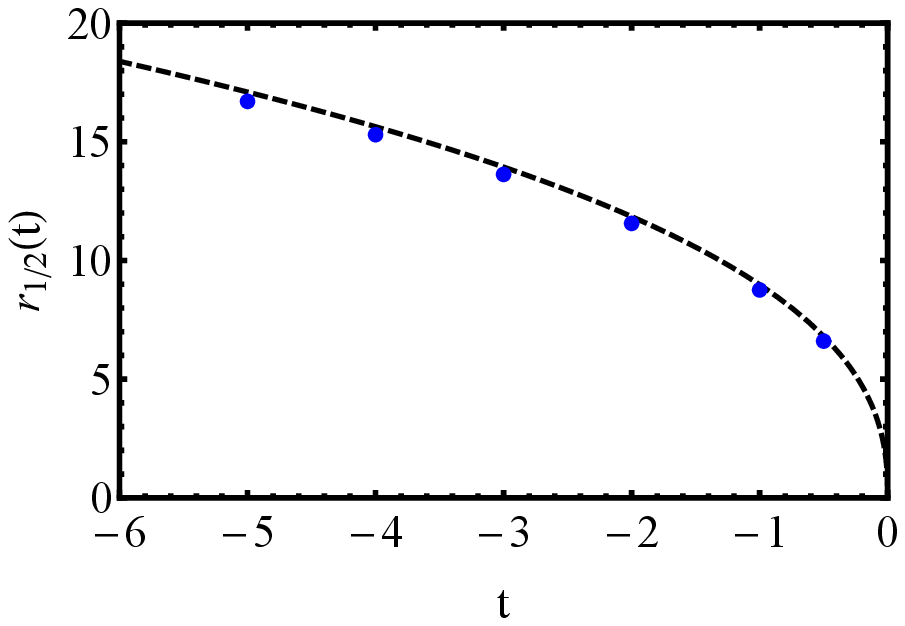}
\caption{Numerically computed $\rho(0,t)$ (left panel, solid line) and $r_{1/2}(t)$ (right panel, symbols) as functions of $t$ for the same parameters as in Fig. \ref{HDsim1}. Dashed lines: theoretical predictions (\ref{HDasympt}).}
\label{HDsim2}
\end{figure}

Figures~\ref{HDsim1} and~\ref{HDsim2} verify the self-similar character of the numerical solution at times that are sufficiently long, but not too close to $t=0$. The left panel of Fig.~\ref{HDsim1} shows the radial profiles of $\rho(r,t)$ at four different moments of time. The right panel depicts the same four radial profiles in rescaled coordinates. Shown is the ratio $\rho(r,t)/\rho(0,t)$ as a function of $r/r_{1/2}(t)$ where, at each time, $r_{{1/2}}(t)$ is the radius where $\rho(r,t)$ is equal to one half of its value at the center, $\rho(0,t)$. The rescaled profiles make a single curve which agrees very well with the theoretical parabolic profile~(\ref{n480}).  Small, but pronounced deviations at the periphery of the ``gas cloud" come from small diffusion effects, unaccounted for by the inviscid hydrodynamic equations~(\ref{n010}) and (\ref{n020}).

Figure~\ref{HDsim2} compares  $\rho(0,t)$ and $r_{1/2}(t)$, found numerically, and their
self-similar asymptotics
\begin{equation}\label{HDasympt}
\rho(0,t) = \frac{3 \ell_0^2}{25} \left(-t\right)^{-6/5} \quad\mbox{and} \quad r_{1/2} (t)=\frac{\ell_0}{\sqrt{2}} \left(-t\right)^{2/5},
\end{equation}
respectively, see Eqs.~(\ref{n420}) and (\ref{n480}). Again, one can see a good agreement.

\section{Implications for directed polymer in random potential}
\label{polymer}

The Hopf-Cole ansatz,
\begin{equation}\label{HC10}
Z(\mathbf{x},t) = \exp\left[\frac{\lambda h(\mathbf{x},t)}{2\nu}\right],
\end{equation}
transforms the KPZ equation (\ref{KPZda}) into a linear diffusion equation with a space-time random potential,
\begin{equation}\label{polymer1}
\partial_{t} Z = \nabla^2 Z + \xi(\mathbf{x},t) Z\,,
\end{equation}
whereas $Z(\mathbf{x},t)$ can be interpreted as the partition function of a continuum directed polymer \cite{KPZ}. This partition function is a random quantity. At $d>2$ one is interested in the probability distribution $\mathcal{F}(Z)$ of $Z$ at steady state, and in its moments $\langle Z^n\rangle$, $n=0,1,2, \dots$. It has been rigorously proven that, at $d>2$, the second moment $\langle Z^2\rangle$ diverges at a finite value $\epsilon=\epsilon_2$ \cite{Comets}. It is generally believed that $\epsilon_2$ is strictly less than the critical point $\epsilon_c$ of the phase transition between the weak and strong coupling. As we show now, the slower-decaying tail (\ref{slowtail}) has important implications in the behavior of the moments $\langle Z^n\rangle$.
Indeed, the steady-state distribution $\mathcal{F}(Z)$ is simply related to our $\mathcal{P}(H)$:
$$
\mathcal{F}(Z) dZ = \mathcal{P}(H) dH.
$$
As a result, the exponential tail (\ref{slowtail}) of $\mathcal{P}(H)$ becomes a \emph{power-law} tail of $\mathcal{F}(Z)$ at $Z\to \infty$:
\begin{equation}\label{PT230}
\mathcal{F}(Z\to \infty) \propto Z^{-\gamma-1}\,,\quad \gamma=\frac{2C_s}{\epsilon}\, .
\end{equation}
It is clear from Eq.~(\ref{PT230})  that the zeroth moment of the steady-state distribution $\mathcal{F}(Z)$ always exists, so the distribution is normalizable to unity. There is, however,  a set of values
\begin{equation}\label{PT260}
    \epsilon_n=\frac{2C_s}{n}\,, \quad n=1,2, \dots\,,
\end{equation}
for which the higher distribution moments $\langle Z^n\rangle$ diverge. These values are ordered as follows:
\begin{equation}\label{PT250}
0=\epsilon_\infty<\dots<\epsilon_{n+1}<\epsilon_n<\dots<\epsilon_1<\epsilon_0
=\infty\, ,
\end{equation}
The set $\{\epsilon_n\}$ depends on $d$ and on the particular form of the noise correlator. It is dominated by the processes on the correlation length $\ell$. There is some uncertainty in interpretation of these results, because the value of the critical point $\epsilon_c$ of the phase transition between the weak and strong coupling is presently unknown. In any case,  the analysis of this section implies that the existence or non-existence of the moments $\langle Z^n\rangle$  is not necessarily related to the phase transition between the weak and strong coupling.

\section{Summary and discussion}
\label{SD}

Nonequilibrium steady states (NESS) of driven, spatially extended stochastic systems have been in the focus of nonequilibrium statistical mechanics for the last two decades. A remarkable progress has been achieved in the analysis of NESS of diffusive lattice gases driven from their boundaries, see Refs. \cite{D07,Jona} for reviews. Here we found another interesting example of a nonequilibrium steady state by considering the one-point height distribution $\mathcal{P}(H)$ emerging when an initially flat interface is driven by a weak noise, as described by the KPZ equation at $d>2$.
The \emph{typical} height fluctuations {correspond to the central part of $\mathcal{P}(H)$. These are Gaussian and belong to the Edwards-Wilkinson universality class. They are completely described by the nonlocal free energy of the interface, Eq.~(\ref{freeenergy}), which makes it redundant to deal with the interface dynamics. In contrast to these,  the distribution tails~(\ref{slowtail}) and (\ref{fasttail}) are non-Gaussian and highly asymmetric. Their activation paths, predicted  by the OFM, are different from the time-reversed relaxation paths, as to be expected far from equilibrium.  The characteristic length scale of the noise, which dominates the contribution to $\mathcal{P}(H)$, is the noise correlation length $\ell$ \cite{KK2008}. The slower-decaying tail (\ref{slowtail}) has  immediate implications for the existence or non-existence of the distribution moments of the partition function $Z$ of the directed polymer.

We hope that our theoretical predictions (\ref{ansatz})-(\ref{fasttail}) will be tested in numerical solutions, at $d=3$, of the KPZ equation with a sufficiently weak short-correlated Gaussian noise. One should measure the (properly shifted) one-point height distribution, including its tails.

It is generally accepted that the transition between the weak and strong couplings in the KPZ equation at $d>2$, as measured by the interface roughness, has a character of a phase transition as a function of $\epsilon$. The interface roughness is an integral quantity, dominated by the typical fluctuations of the interface. The one-point height distribution provides a more detailed characterization of the fluctuations. For finite $\epsilon$, which are below the phase transition,
the distribution of \emph{typical} fluctuations may change compared with the prediction of Eq.~(\ref{variance}). The far tails, however,
should be still describable by Eqs.~(\ref{slowtail}) and (\ref{fasttail}).  It would be interesting to see, in an ``infinite" system, how the stationary height distribution for the weak coupling, that we studied here, gives way (via a phase transition) to a non-stationary height distribution, which exhibits a universal scaling behavior, as observed  in numerical simulations of the strong-coupling regime \cite{HH2013,Braziliangroup}.

For $d=1$, an initially weak noise becomes effectively strong at long times. Still, it was conjectured in Ref. \cite{MKV}, that the far-tail asymptotics, predicted by the OFM (which formally is a ``weak-noise" theory), continue to hold at arbitrarily long times. For the sharp-wedge initial interface this conjecture was recently proved both for the  $\lambda H>0$ tail \cite{Majumdar}, and for the  $\lambda H<0$ tail \cite{SMP, CorwinGhosal,KLD2018,Corwinetal2018}. Based on this analogy, we argue that, at $d>2$, the time-independent asymptotics (\ref{slowtail}) and (\ref{fasttail}) persist, at any finite time and at sufficiently large $|H|$, in the strong-coupling regime as well.

Finally, what happens in the marginal case $d=2$? Here $\mathcal{P}_t(H)$ does \emph{not} reach a steady state: at long times it continues to change with time, albeit logarithmically slowly. Similarly to the case of $d<2$, an initially weak noise ultimately becomes strong at $d=2$, but this happens only at times which, for a small correlation length $\ell$, are exponentially long \cite{NattermannTang}. In this special case, the OFM problem is technically more involved and demands a separate consideration.

\section*{Acknowledgments}

We are very grateful to Joachim Krug for illuminating discussions and to Herbert Spohn for a valuable comment.  B. M.
acknowledges support from the Israel Science Foundation (Grant No. 807/16)
and from the University of Cologne through the Center of Excellence
``Quantum Matter and Materials."

\begin{appendices}


\subsection*{Appendix. Numerical algorithms}
\label{Appendix:numerics}

\renewcommand{\theequation}{A\arabic{equation}}
\setcounter{equation}{0}

Here we will briefly describe the two numerical algorithms used in this work.

\subsubsection{Solving the time-dependent OFM equations}

Assuming $d=3$, spherical symmetry and the Gaussian correlator (\ref{correlator}) with $\ell=1$,
we can integrate  Eq.~(\ref{pi}) over the spherical angles. The result can be written as
\begin{equation}\label{pisphrc}
\pi(r,t)=\frac{1}{\sqrt{\pi}}\int_{0}^{\infty}dr'(r')^{2}\rho(r't)\frac{e^{-(r-r')^{2}}-e^{-(r+r')^{2}}}{r r'}.
\end{equation}
It is convenient to go over from $\rho(r,t)$ to a new variable $w(r,t)$ which eliminates the delta-singularity in the
boundary condition~(\ref{ptimezero}):
\begin{equation}\label{wdefinition}
w(r,t)=\frac{\Lambda}{4\pi}\int_{0}^{r}dr'(r')^{2}\rho(r't).
\end{equation}
The problem, described by Eqs.(\ref{heqd})-(\ref{pi}) with the boundary condition~(\ref{ptimezero}), becomes
\begin{eqnarray}
\partial_th &=& \frac{1}{r^2} \partial_{r}(r^{2}\partial_{r}h) -\frac{1}{2} (\partial_{r} h)^{2}+\frac{\Lambda}{4\pi^{3/2}}\int_{0}^{\infty}dr'\partial_{r'}w(r',t)\,\frac{e^{-(r-r')^{2}}-e^{-(r+r')^{2}}}{r r'} \, ,
\label{hnew}\\
\partial_t w & = & -\partial_{r}^2 w+\frac{2}{r}\partial_{r}w-\partial_{r}w\,\partial_{r}h \, ,
\label{weqn}\\
w(0,t) &=&0,\quad\partial_{r}w(\infty,t)=0,\quad w(r,0)=1/2.
\label{wb.c}
\end{eqnarray}
The Hopf-Cole ansatz $q(r,t)=\exp[-h(r,t)/2]$ turns Eq.~(\ref{hnew}) into a  linear equation for $q(r,t)$:
\begin{equation}\label{qeqn}
\partial_{t}q= \partial_{rr}q+\frac{2}{r} \partial_{r} q +\frac{\Lambda}{8\pi^{3/2}}q\int_{0}^{\infty}dr'\partial_{r'}w(r',t)\,
\frac{e^{-(r-r')^{2}}-e^{-(r+r')^{2}}}{r r'} .
\end{equation}
Once the solution of Eqs.~(\ref{weqn}) and (\ref{qeqn}) is obtained, the action is
\begin{equation}\label{wactn}
s=2\sqrt{\pi}\int_{-\infty}^{0}dt\int_{0}^{\infty}dr'\partial_{r'}w(r',t)\int_{0}^{\infty}dr\partial_{r}w(r,t)
\,\frac{e^{-(r-r')^{2}}-e^{-(r+r')^{2}}}{r r'}.
\end{equation}
We solved finite in time and space versions of discretized Eqs.~(\ref{weqn})-(\ref{qeqn}) numerically by using the
Chernykh-Stepanov back-and-forth iteration algorithm \cite{CS}.

\subsubsection{Solving the integro-differential equation~(\ref{H234})}
We solved the spherically-symmetric version of Eq.~(\ref{H234}) for the Gaussian correlator (\ref{correlator}) with $\ell=1$:
\begin{equation}\label{Z3D}
\frac{1}{r^2} \frac{d}{dr}\left(r^{2}\frac{dZ}{dr}\right)=Z(r)\left[1-\int_{0}^{\infty}dr'(r')^{2}Z^{2}(r')\kappa_{c}(r,r')\right],\quad
\kappa_{c}(r,r')=\sqrt{\frac{2}{\pi c}}\frac{e^{-\frac{2}{c}(r-r')^{2}}-e^{-\frac{2}{c}(r+r')^{2}}}{r r'}.
\end{equation}
The solution should be everywhere positive and satisfy the boundary conditions
\begin{equation}\label{Zbc}
\frac{dZ(0)}{dr}=0,\quad Z(\infty)=0.
\end{equation}
Consider a modified and truncated version of the problem~(\ref{Z3D}) and (\ref{Zbc}), where the infinity at the upper integration limit  is replaced by some $X>0$, and  the second condition in Eq.~(\ref{Zbc}) is replaced by a different condition:
\begin{eqnarray}
 \frac{1}{r^2} \frac{d}{dr}\left(r^{2}\frac{dz}{dr}\right)&=&  z(r)\left[1-\int_{0}^{X}dr'(r')^{2}z^{2}(r')\kappa_{c}(r,r')\right], \label{Z3Dm} \\
  \frac{dz(0,X)}{dr} &=& 0, \label{Zbc1}\\
  z(0,X)&=& z_0 . \label{Zbc2}
\end{eqnarray}
Let us call its solution $z=z(r,X,z_0)$.   We solve this problem~(\ref{Z3Dm})-(\ref{Zbc2}) iteratively. We start  from a very small $X$, $X\ll 1/\sqrt{c}$, and approximate the problem on three grid points:  $r=0$, $r=X/2$, and $r=X$. $z(0)$ is given by Eq.~(\ref{Zbc2}), whereas Eqs.~(\ref{Z3Dm}) and (\ref{Zbc1}) give two algebraic equations: a linear one and a non-linear one, which can be solved. The quantity $X/2$ for this first iteration serves as the mesh size in the subsequent iterations, and we will call it $\delta r$.

In the next step we increase $X$ by adding $\delta r$ to the previous value of $X$. Now we have four grid points. $z(0)$ is still given by Eq.~(\ref{Zbc2}). Equation~(\ref{Zbc1}) still gives one (linear) algebraic equation,   whereas Eqs.~(\ref{Z3Dm}) now gives two nonlinear algebraic equations at the points $r=\delta r$ and $r=2 \delta r$. Importantly, the values of $z(r=\delta r)$ and $z(r=2\delta r)$ from the previous step serve as initial guesses in the iterative solution of the algebraic equations.

Continuing increasing $X$ by adding $\delta r$ in this way we can, in principle, solve the modified problem for large $X$.
This solution, however, would not satisfy the second boundary condition (\ref{Zbc}), as it either becomes negative at some $r$ or increases with $r$, which is inadmissible. Therefore, we change the value of $z_0$ in Eq.~(\ref{Zbc2}) and repeat the iterations. This ``shooting" procedure can be conveniently organized by bisections, and it continues until the solution converges with desired accuracy.

\end{appendices}

\end{document}